\documentclass[11pt]{article}
\usepackage{amssymb,amsmath,amsfonts}
\usepackage{graphicx}
\usepackage{graphics}
\usepackage{eepic,epsfig}

\@addtoreset{equation}{section}

\makeatother

\begin{document}

\title{Fermionic condensate and the Casimir effect\\
in cosmic string spacetime }
\author{A. Kh. Grigoryan$^{1}$\thanks{%
E-mail: ashot.gr@gmail.com},\thinspace\, A. R. Mkrtchyan$^{1,2}$\thanks{%
E-mail: malpic@sci.am}, \thinspace\, A. A. Saharian$^{1,3}$\thanks{%
E-mail: saharian@ysu.am}, \vspace{0.3cm} \\
\textit{$^1$Institute of Applied Problems in Physics NAS RA,}\\
\textit{25 Nersessian Street, 0014 Yerevan, Armenia}\vspace{0.3cm}\\
\textit{$^2$Tomsk Polytechnic University, 30 Lenin Avenue, 634050 Tomsk,
Russia}\vspace{0.3cm}\\
\textit{$^3$Department of Physics, Yerevan State University,}\\
\textit{1 Alex Manoogian Street, 0025 Yerevan, Armenia}}
\date{}
\maketitle

\begin{abstract}
We investigate combined effects of nontrivial topology, induced by a cosmic
string, and boundaries on the fermionic condensate and the vacuum
expectation value (VEV) of the energy-momentum tensor for a massive
fermionic field. As geometry of boundaries we consider two plates
perpendicular to the string axis on which the field is constrained by the
MIT bag boundary condition. By using the Abel-Plana type summation formula,
the VEVs in the region between the plates are decomposed into the
boundary-free and boundary-induced contributions for general case of the
planar angle deficit. The boundary-induced parts in both the fermionic
condensate and the energy-momentum tensor vanish on the cosmic string.
Fermionic condensate is positive near the string and negative al large
distances, whereas the vacuum energy density is negative everywhere. The
radial stress is equal to the energy density. For a massless field, the
boundary-induced contribution in the VEV of the energy-momentum tensor is
different from zero in the region between the plates only and it does not
depend on the coordinate along the string axis. In the region between the
plates and at large distances from the string, the decay of the topological
part is exponential for both massive and massless fields. This behavior is
in contrast to that for the VEV of the energy-momentum tensor in the
boundary-free geometry with the power law decay for a massless field. The
vacuum pressure on the plates is inhomogeneous and vanishes at the location
of the string. The corresponding Casimir forces are attractive.
\end{abstract}

\bigskip


\bigskip

\section{Introduction}

In a large number of field theoretical problems one needs to consider
quantum fields in backgrounds with boundaries. The boundary conditions
imposed on the field operator modify the spectrum of the zero-point
fluctuations affecting the vacuum expectation values (VEVs) of physical
observables. Among the most interesting manifestations of this influence is
the Casimir effect (for reviews see \cite{Most97}). In 1948, Casimir showed
that two neutral parallel conducting plates placed in the vacuum attract
each other by the force inversely proportional to the fourth power of the
distance between them. The appearance of the force was explained as a
consequence of the change in the spectrum for the vacuum fluctuations of the
electromagnetic field caused by the presence of the plates. Since the
Casimir's original paper, the vacuum forces acting on the boundaries
confining quantum fields have been investigated both theoretically and
experimentally for other boundary geometries and also for scalar and
fermionic fields. Similar effects arise in models with nontrivial spatial
topology. In these models the boundary conditions imposed on the fields have
topological origin.

Among the most interesting directions in the theory of the Casimir effect is
its dependence on the geometry of the background spacetime. In particular,
motivated by the possibility for the Casimir forces to serve as a
stabilization mechanism for the radion field in braneworld models, the
investigation of the Casimir effect in anti-de Sitter spacetime has
attracted a great deal of attention (see, e.g., references given in \cite%
{Beze15AdS}). In the present paper we consider a problem with two types of
the sources for the fermionic vacuum polarization. The first one is
topological and is induced by the presence of straight cosmic string and the
second one is induced by two parallel conical boundaries. The cosmic strings
are among the most interesting topological defects resulting from the
symmetry braking phase transitions in grand unified theories \cite{Vile94}.
In particular, they may have been created in the early Universe and provide
an important link between particle physics and cosmology. A variant of the
formation mechanism for cosmic strings has been proposed in models of brane
inflation \cite{Sara02}. Although the recent observational data on the
cosmic microwave background radiation (CMB) have ruled out cosmic strings as
the primary source for primordial density perturbations, they are still
candidates for the generation of a number of interesting physical effects.
The latter include the gravitational lensing, the creation of small
non-Gaussianities in the CMB, the generation of gravitational waves,
high-energy cosmic rays, and gamma ray bursts. The conical space appears
also as an effective background geometry in the long-wavelength description
of condensed matter systems such as crystals, liquid crystals and quantum
liquids \cite{Nels02}. Among condensed matter realizations of the fermionic
model in the conical bulk are graphene nanocones. The latter are obtained
from planar graphene sheet if one or more sectors with the angle $\pi /3$
are excised and the remainder is joined \cite{Kris97}. The long-wavelength
excitations of the electronic subsystem in graphene are described by
Dirac-like theory with the Fermi velocity playing the role of the speed of
light. In the case of graphene nanocones the background geometry for this
theory is conical.

In the present paper we are interested in the influence of the nontrivial
topology due to the cosmic string on the fermionic Casimir densities and the
Casimir force in the geometry of two parallel plates with the MIT bag
boundary condition. In order to have an exactly solvable problem, we will
consider the idealized model for the straight cosmic string with flat
spacetime everywhere except on the string where the curvature tensor has
delta-type singularity. In this simplified model the exterior spacetime has
a conical structure with the planar angle deficit related to the linear mass
density. Among the interesting physical effects of the corresponding
nontrivial topology, widely discussed in the literature, is the polarization
of the vacuum for quantum fields (see, for instance, references given in
\cite{Bell14}). Moreover, combined effects of the topology and boundaries in
the geometry of the cosmic string have been considered as well. The analysis
of the boundary-induced quantum effects in the cosmic string spacetime have
been developed for scalar \cite{Beze06sc,Nest11}, fermionic \cite%
{Beze08,Beze10}, and electromagnetic fields \cite{Nest11,Brev95,Beze07} in
the geometry of a coaxial cylindrical boundary. The Casimir force for a
massless scalar fields subject to Dirichlet and Neumann boundary conditions
in the setting of the conical piston has been investigated in \cite{Fucc11}.
The Casimir densities for scalar and electromagnetic fields induced by flat
boundaries perpendicular to the string were considered in \cite%
{Beze11,Beze12,Beze13}.

As local characteristics of the fermionic vacuum we will consider the
fermionic condensate and the VEV of the energy-momentum tensor. The notion
of the vacuum has a global nature and, as a consequence of this, these VEVs
carry an important information about the global properties of the background
spacetime. The fermionic condensate plays a crucial role in the models of
chiral symmetry breaking and in considerations of the stability of the
fermionic vacuum \cite{Higa91,Inag97}. In addition to describing the local
properties of the vacuum, the VEV of the energy-momentum tensor acts as the
source in semiclassical Einstein equations and is required in modelling the
self-consistent dynamics involving the gravitational field.

The paper is organized as follows. In the next section we specify the bulk
and boundary geometries and the boundary conditions imposed on the fermion
field. The complete set of the positive- and negative-energy wavefunctions
is presented. By making use of these modes, in section \ref{sec:FC} we
evaluate the fermionic condensate in the region between the plates. Various
asymptotic limits of the general expression are discussed. Similar
considerations for the VEV of the energy-momentum tensor are presented in
section \ref{sec:EMT}. The Casimir forces acting on the plates are
investigated in section \ref{sec:Force}. The main results are summarized in
section \ref{sec:Conc}. Some details in the transformations of the
components for the vacuum energy-momentum tensor are presented in Appendix %
\ref{sec:Appendix}.

\section{Geometry of the problem and the mode functions}

\label{sec:Modes}

We start with the description of the bulk and boundary geometries under
consideration. The background geometry is generated by an infinitely long
straight cosmic string along the $z$ axis. In the cylindrical coordinates $%
x^{\mu }=(t,r,\phi ,z)$ the corresponding metric tensor reads $g_{\mu \nu }=%
\mathrm{diag}(1,-1,-r^{2},-1)$. The difference from the Minkowskian
spacetime comes from the planar angle deficit $2\pi -\phi _{0}$ for the
azimuthal angle $\phi $, $0\leqslant \phi \leqslant \phi _{0}$. In the
discussion below we will also use the parameter $q=2\pi /\phi _{0}$. In the
weak-field approximation, one has the relation $2\pi -\phi _{0}=8\pi G\mu
_{0}$, where $\mu _{0}$ is the linear mass density on the string and $G$ is
the Newton gravitational constant.

For points outside the string core, the local geometry generated by the
cosmic string is flat and coincides with that for the Minkowski spacetime.
However, these two spacetimes have different global properties. The
nontrivial topology coming from the cosmic string induces shifts in the VEVs
of physical observables for quantum fields. Here we are interested in the
influence of topology on the Casimir effect for a quantum fermionic field $%
\psi (x)$ in the geometry of two parallel plates perpendicular to the string
axis. The field operator obeys the Dirac equation
\begin{equation}
\left( i\gamma ^{\mu }\nabla _{\mu }-m\right) \psi (x)=0,  \label{Direq}
\end{equation}%
with the covariant derivative $\nabla _{\mu }=\partial _{\mu }+\Gamma _{\mu
} $ and with $\Gamma _{\mu }$ being the spin connection. In cylindrical
coordinates with a planar angle deficit, for the Dirac $4\times 4$ matrices $%
\gamma ^{\mu }$ we will take the representation%
\begin{equation}
\gamma ^{0}=\left(
\begin{array}{cc}
1 & 0 \\
0 & -1%
\end{array}%
\right) ,\;\gamma ^{l}=\left(
\begin{array}{cc}
0 & \beta ^{l} \\
-\beta ^{l} & 0%
\end{array}%
\right) ,  \label{gam0l}
\end{equation}%
where $\beta ^{l}$, $l=1,2,3$, are the Pauli matrices in the same
coordinates. The latter are presented as
\begin{equation}
\beta ^{l}=(i/r)^{l-1}\left(
\begin{array}{cc}
0 & (-1)^{l-1}e^{-iq\phi } \\
e^{iq\phi } & 0%
\end{array}%
\right) ,\;\beta ^{3}=\left(
\begin{array}{cc}
1 & 0 \\
0 & -1%
\end{array}%
\right) ,  \label{Pauli}
\end{equation}%
with $l=1,2$.

We assume the presence of two boundaries located at $z=0$ and $z=a$ on which
the field operator is constrained by the MIT bag boundary conditions:
\begin{equation}
\left( 1+in_{\mu }\gamma ^{\mu }\right) \psi (x)=0\ ,\quad z=0,a,
\label{BagCond}
\end{equation}%
where $n_{\mu }$ is the outward-pointing normal to the boundary with respect
to the region under consideration. In the discussion below we will consider
the region $0\leqslant z\leqslant a$ with $n_{\mu }=-\delta _{\mu 1}$ and $%
n_{\mu }=\delta _{\mu 1}$ for the boundaries at $z=0$ and $z=a$
respectively. In the regions $z\leqslant 0$ and $z\geqslant a$, the VEVs are
obtained by the limiting transitions and they are the same as in the
corresponding problem with a single boundary, discussed in \cite{Beze13}.
Note that the boundaries $z=0,a$ are two-dimensional cones with the angle
deficit $2\pi -\phi _{0}$. The models with boundary conditions induced by
the compactification of the cosmic string along its axis have been
considered in \cite{Beze12b}.

We are interested in the investigation of the combined effects of topology
and boundaries on the fermion condensate and on the VEV\ of the
energy-momentum tensor. These quantities are expressed in the form of the
sums over complete set of fermionic modes obeying the boundary conditions (%
\ref{BagCond}). So, our first step will be the determination of the complete
set of the fermionic wavefunctions in the region between the boundaries. In
\cite{Beze13} it has been shown that, in the geometry of a single plate at $%
z=0$, the positive-energy mode functions obeying the boundary condition (\ref%
{BagCond}) on that plate are presented as
\begin{equation}
\psi _{\sigma }^{(+)}=C_{\sigma }^{(+)}e^{i\left( qj\phi -\omega t\right)
}\left(
\begin{array}{c}
f_{+}(z)J_{\beta }(\lambda r)e^{-iq\phi /2} \\
i\frac{\epsilon _{j}s}{k}g_{+}(z)J_{\beta +\epsilon _{j}}(\lambda
r)e^{iq\phi /2} \\
\frac{1}{k}g_{-}(z)J_{\beta }(\lambda r)e^{-iq\phi /2} \\
-i\epsilon _{j}sf_{-}(z)J_{\beta +\epsilon _{j}}(\lambda r)e^{iq\phi /2}%
\end{array}%
\right) ,  \label{psiPos}
\end{equation}%
where $J_{\beta }(x)$ is the Bessel function, $0\leqslant \lambda <0$, $%
\epsilon _{j}=1$ for $j\geqslant 0$ and $\epsilon _{j}=-1$ for $j<0$ with $%
j=\pm 1/2,\pm 3/2,\ldots $, and
\begin{equation}
\beta =q|j|-\epsilon _{j}/2,\;\omega =\sqrt{\lambda ^{2}+k^{2}+m^{2}}.
\label{beta}
\end{equation}%
In (\ref{psiPos}) we have defined the functions%
\begin{eqnarray}
f_{\pm }(z) &=&e^{ikz}\pm i\kappa _{s}e^{-ikz},  \notag \\
g_{\pm }(z) &=&\left( \omega \pm m\right) f_{\pm }(z)+s\lambda f_{\mp }(z),
\label{fz}
\end{eqnarray}%
with $k>0$ and%
\begin{equation}
\kappa _{s}=\frac{\omega +s\lambda }{k-im},\;s=-1,1.  \label{kapas}
\end{equation}%
In (\ref{psiPos}), $\sigma $ stands for the set of quantum numbers
specifying the mode functions (see below).

Now we should impose on the modes (\ref{psiPos}) the boundary condition (\ref%
{BagCond}) at $z=a$. From the latter it follows that the eigenvalues of the
quantum number $k$ are determined by the transcendental equation%
\begin{equation}
e^{2ika}=\frac{m-ik}{m+ik}.  \label{EigEq}
\end{equation}%
The latter can also be written in the form%
\begin{equation}
ma\sin (ka)+ka\cos (ka)=0.  \label{EigEq2}
\end{equation}%
This equation has an infinite number of positive solutions with respect to $%
ka$. We will denote them by $x_{n}=ka$, $n=1,2,\ldots $, $x_{n+1}>x_{n}$.
For the eigenvalues of $k$ one has $k=k_{n}=x_{n}/a$.

For the negative-energy mode functions, in a similar way, one has the
expression%
\begin{equation}
\psi _{\sigma }^{(-)}=C_{\sigma }^{(-)}e^{i\left( qj\phi +\omega t\right)
}\left(
\begin{array}{c}
-\frac{1}{k}g_{-}(z)J_{\beta }(\lambda r)e^{-iq\phi /2} \\
i\epsilon _{j}sf_{-}(z)J_{\beta +\epsilon _{j}}(\lambda r)e^{iq\phi /2} \\
f_{+}(z)J_{\beta }(\lambda r)e^{-iq\phi /2} \\
i\frac{\epsilon _{j}s}{k}g_{+}(z)J_{\beta +\epsilon _{j}}(\lambda
r)e^{iq\phi /2}%
\end{array}%
\right) ,  \label{psiNeg}
\end{equation}%
with the same notations as in (\ref{psiPos}) and with the same equation (\ref%
{EigEq2}) for the eigenvalues of $ka$. It can be explicitly checked that the
positive- and negative-energy modes (\ref{psiPos}) and (\ref{psiNeg}) are
orthogonal. The four-spinors $\psi _{\sigma }^{(\pm )}$ are eigenfunctions
of the projection of the total angular momentum along the cosmic string:%
\begin{equation}
\hat{J}_{3}\psi _{\sigma }^{(\pm )}=\left( -i\partial _{\phi }+i\frac{q}{2}%
r\gamma ^{1}\gamma ^{2}\right) \psi _{\sigma }^{(\pm )}=qj\psi _{\sigma
}^{(\pm )}.  \label{J3psi}
\end{equation}%
Now, the set of quantum numbers for the modes are specified by $\sigma
=(\lambda ,j,n,s)$.

The coefficients $C_{\sigma }^{(\pm )}$ in (\ref{psiPos}) and (\ref{psiNeg})
are found from the orthonormality conditions%
\begin{equation}
\int d^{3}x\sqrt{|g|}\psi _{\sigma }^{(\pm )\dagger }\psi _{\sigma ^{\prime
}}^{(\pm )}=\delta (\lambda -\lambda ^{\prime })\delta _{nn^{\prime }}\delta
_{j^{\prime }j}\delta _{ss^{\prime }}\ ,  \label{Norm}
\end{equation}%
where the dagger denotes Hermitian conjugation and the integration goes over
the region between the plates. By taking into account that for the Bessel
function one has
\begin{equation}
\int_{0}^{\infty }dr\,rJ_{\beta }(\lambda r)J_{\beta }(\lambda ^{\prime }r)=%
\frac{1}{\lambda }\delta (\lambda -\lambda ^{\prime }),  \label{IntRel}
\end{equation}%
we get%
\begin{equation}
\left\vert C_{\sigma }^{(\pm )}\right\vert ^{-2}=\frac{\phi _{0}}{\lambda }%
\frac{8\omega \left( \omega +s\lambda \right) a}{k^{2}}\left[ 1-\frac{\sin
(2ka)}{2ka}\right] ,  \label{Cnorm}
\end{equation}%
with $k=x_{n}/a$.

\section{Fermionic condensate}

\label{sec:FC}

We start our investigation of the VEVs by the fermion condensate, defined as
the expectation value $\langle 0|\bar{\psi}\psi |0\rangle \equiv \langle
\bar{\psi}\psi \rangle $, with $|0\rangle $ being the vacuum state and $\bar{%
\psi}=\psi ^{\dag }\gamma ^{0}$ is the Dirac adjoint. Having the complete
set of mode functions, the fermion condensate is evaluated by using the mode
sum
\begin{equation}
\langle \bar{\psi}\psi \rangle =\sum_{s=\pm 1}\sum_{j=\pm 1/2,\ldots
}\sum_{n=1}^{\infty }\int_{0}^{\infty }d\lambda \,\bar{\psi}_{\sigma
}^{(-)}\psi _{\sigma }^{(-)}.  \label{FC}
\end{equation}%
We assume that some regularization scheme is used to make the expression in
the right-hand side finite (for example, a cutoff function is introduced or
the arguments of the operators in the product are shifted). The choice of
the specific scheme is not essential in the further discussion and we will
not display it explicitly.

By making use of the expression (\ref{psiNeg}) for the mode functions and
the relations
\begin{equation}
\sum_{j=\pm 1/2,\ldots }J_{\beta }^{2}(\lambda r)=\sum_{j=\pm 1/2,\ldots
}J_{\beta +\epsilon _{j}}^{2}(\lambda r)=\sum_{j}[J_{qj-1/2}^{2}(\lambda
r)+J_{qj+1/2}^{2}(\lambda r)],  \label{RelBes}
\end{equation}%
the fermionic condensate is presented in the form%
\begin{equation}
\langle \bar{\psi}\psi \rangle =-\frac{qm}{2\pi a}\sum_{j}\int_{0}^{\infty
}d\lambda \,\lambda \sum_{n=1}^{\infty }\frac{J_{qj-1/2}^{2}(\lambda
r)+J_{qj+1/2}^{2}(\lambda r)}{\omega \left[ 1-\sin (2x_{n})/(2x_{n})\right] }%
h(k_{n},z).  \label{FCin}
\end{equation}%
Here, $\omega =\sqrt{k_{n}^{2}+\lambda ^{2}+m^{2}}$,
\begin{equation}
h(k,z)=2-\sum_{\eta =\pm 1}\left( 1+\eta ik/m\right) e^{2\eta ikz},
\label{hkz}
\end{equation}%
and $\sum_{j}$ stands for the summation over $j=1/2,3/2,\cdots $.

By using the integral representation%
\begin{equation}
\omega =(2/\sqrt{\pi })\int_{0}^{\infty }ds\,e^{-\omega ^{2}s^{2}},
\label{IntRep}
\end{equation}%
the integral over $\lambda $ in (\ref{FCin}) is evaluated with the help of
the formula \cite{Prud86}%
\begin{equation}
\int_{0}^{\infty }d\lambda \,\lambda J_{qj\pm 1/2}^{2}(\lambda r)e^{-\lambda
^{2}s^{2}}=\frac{I_{qj\pm 1/2}(r^{2}/2s^{2})}{2s^{2}e^{r^{2}/2s^{2}}},
\label{Jint}
\end{equation}%
where $I_{\nu }(x)$ is the modified Bessel function. As a result, the
fermionic condensate is presented in the form%
\begin{eqnarray}
\langle \bar{\psi}\psi \rangle  &=&-\frac{qm}{(2\pi )^{3/2}ar}%
\sum_{n=1}^{\infty }\frac{h(x_{n}/a,z)}{1-\sin (2x_{n})/(2x_{n})}  \notag \\
&&\times \int_{0}^{\infty }dy\,y^{-1/2}e^{-(x_{n}^{2}/a^{2}+m^{2})r^{2}/2y-y}%
\mathcal{I}(q,y),  \label{FC1}
\end{eqnarray}%
with the notation%
\begin{equation}
\mathcal{I}(q,y)=\sum_{j}\left[ I_{qj-1/2}(y)+I_{qj+1/2}(y)\right] .
\label{Ical}
\end{equation}

As the next step in the evaluation of the fermionic condensate, for the
series (\ref{Ical}) we use the representation \cite{Beze10}%
\begin{eqnarray}
\mathcal{I}(q,x) &=&\frac{2}{q}\sideset{}{'}{\sum}_{l=0}^{p}(-1)^{l}\cos
\left( \pi l/q\right) e^{x\cos (2\pi l/q)}  \notag \\
&&+\frac{2}{\pi }\cos \left( q\pi /2\right) \int_{0}^{\infty }du\frac{\sinh
\left( qu/2\right) \sinh \left( u/2\right) }{\cosh (qu)-\cos (q\pi )}%
e^{-x\cosh u},  \label{RepIcal}
\end{eqnarray}%
where $p$ is the integer part of $q/2$, $p=[q/2]$. In (\ref{RepIcal}), the
prime on the sign of the sum means that the term $l=0$ and the term $l=p$
for even values of $q$ should be taken with the coefficients 1/2. In the
absence of the cosmic string one has $q=1$ and we get $\mathcal{I}%
(1,y)=e^{y} $. From here it follows that the contribution of the term $l=0$
in (\ref{RepIcal}) to $\langle \bar{\psi}\psi \rangle $ coincides with the
fermionic condensate in the region between two plates in the Minkowski
spacetime. Substituting (\ref{RepIcal}) into (\ref{FC1}), the integral over $%
y$ is explicitly evaluated and one finds the following representation%
\begin{eqnarray}
\langle \bar{\psi}\psi \rangle &=&-\frac{m}{2\pi ar}\sum_{n=1}^{\infty }%
\frac{h(x_{n}/a,z)}{1-\sin (2x_{n})/(2x_{n})}\left[ \sideset{}{'}{\sum}%
_{l=0}^{p}(-1)^{l}\frac{c_{l}}{s_{l}}e^{-2rs_{l}\sqrt{k_{n}^{2}+m^{2}}%
}\right.  \notag \\
&&\left. +\frac{2q}{\pi }\int_{0}^{\infty }dx\frac{\cos \left( q\pi
/2\right) \sinh \left( qx\right) \sinh x}{\cosh (2qx)-\cos (q\pi )}\frac{%
e^{-2r\sqrt{k_{n}^{2}+m^{2}}\cosh x}}{\cosh x}\right] ,  \label{FC2}
\end{eqnarray}%
with the notations%
\begin{equation}
s_{l}=\sin (\pi l/q),\;c_{l}=\cos \left( \pi l/q\right) .  \label{sl}
\end{equation}

For the general case of a massive field, the eigenvalues $k_{n}$ in (\ref%
{FC2}) are given implicitly, as roots of (\ref{EigEq2}), and this form for
the fermionic condensate is not convenient for the further investigation of
its properties. For the summation over these eigenvalues we use the formula%
\begin{equation}
\sum_{n=1}^{\infty }\frac{\pi f(x_{n})}{1-\sin (2x_{n})/(2x_{n})}=-\frac{\pi
maf(0)}{2(ma+1)}+\int_{0}^{\infty }dx\,f(x)-i\int_{0}^{\infty }dx\frac{%
f(ix)-f(-ix)}{\frac{x+ma}{x-ma}e^{2x}+1},  \label{Sumxn}
\end{equation}%
that is a consequence of the generalized Abel-Plana summation formula \cite%
{SahaBook}. For the series in (\ref{FC2}) the corresponding function is
given by the expression
\begin{equation}
f(x)=h(x/a,z)e^{-2br\sqrt{x^{2}/a^{2}+m^{2}}}.  \label{fxFC}
\end{equation}%
For this function $f(0)=0$ and the first term in the right-hand side of (\ref%
{Sumxn}) is absent. The first integral in (\ref{Sumxn}) is evaluated by
using the formula%
\begin{equation}
\int_{m}^{\infty }dx\sin (2y\sqrt{x^{2}-m^{2}})e^{-2zx}=2ym^{2}f_{1}(2m\sqrt{%
z^{2}+y^{2}}),  \label{form1}
\end{equation}%
with the notation
\begin{equation}
f_{\nu }(x)=K_{\nu }(x)/x^{\nu },  \label{fnu}
\end{equation}%
and the relation
\begin{equation}
\int_{0}^{\infty }dx\,\sum_{\eta =\pm 1}\left( m+\eta ix\right) e^{2\eta
ixz}e^{-2y\sqrt{x^{2}+m^{2}}}=\left( 2m+\partial _{z}\right)
\int_{m}^{\infty }dx\,\sin (2y\sqrt{x^{2}-m^{2}})e^{-2xz}.  \label{form2}
\end{equation}%
In deriving (\ref{form2}) we have rotated the integration contour of the
left-hand side in the complex $x$-plane by the angle $\pi /2$ for $\eta =1$
and by $-\pi /2$ for $\eta =-1$. As a result one finds the representation%
\begin{equation}
\frac{\pi }{a}\sum_{n=1}^{\infty }\frac{h(x_{n}/a,z)e^{-2y\sqrt{%
x_{n}^{2}/a^{2}+m^{2}}}}{1-\sin (2x_{n})/(2x_{n})}=4m^{2}y\left[
f_{1}(2my)-F(2mz,2my)\right] -\frac{y}{m}B(z,y),  \label{SerSum}
\end{equation}%
where we have defined the functions%
\begin{equation}
F(x,y)=f_{1}(\sqrt{x^{2}+y^{2}})-xf_{2}(\sqrt{x^{2}+y^{2}}),  \label{Fxy}
\end{equation}%
and%
\begin{equation}
B(z,y)=\frac{2}{y}\int_{m}^{\infty }dx\frac{\sin (2y\sqrt{x^{2}-m^{2}})}{%
\frac{x+m}{x-m}e^{2ax}+1}\left[ 2m-\sum_{\eta =\pm 1}(m+\eta x)e^{2\eta zx}%
\right] .  \label{FzyFC}
\end{equation}%
Here, the relation $f_{\nu }^{\prime }(x)=-xf_{\nu +1}(x)$ is used for the
function (\ref{fnu}).

Substituting (\ref{SerSum}) into (\ref{FC2}), the fermionic condensate is
decomposed into three contributions. The first one comes from the term with $%
f_{1}(2my)$ in (\ref{SerSum}). This contribution does not depend on $z$ and
on $a$ and corresponds to the fermionic condensate in the boundary-free
cosmic string spacetime. For points away from the boundaries, the only
divergence in the fermionic condensate is contained in the $l=0$ term of
this contribution. The latter coincides with the fermionic condensate in
boundary-free Minkowski spacetime. The renormalization is reduced to
omitting this term. As a result, the renormalized fermionic condensate in
the boundary-free cosmic string geometry is presented as%
\begin{eqnarray}
\langle \bar{\psi}\psi \rangle _{\mathrm{s}} &=&-\frac{2m^{3}}{\pi ^{2}}%
\bigg[\sum_{l=1}^{p}(-1)^{l}c_{l}f_{1}(2mrs_{l})+\frac{2q}{\pi }\cos \left(
\frac{q\pi }{2}\right)  \notag \\
&&\times \int_{0}^{\infty }dx\frac{\sinh \left( qx\right) \sinh
(x)f_{1}(2mr\cosh x)}{\cosh (2qx)-\cos (q\pi )}\bigg].  \label{FCsren}
\end{eqnarray}%
This expression has been previously derived in \cite{Beze13} (for the
generalization in the geometry of cosmic string with a magnetic flux see
\cite{Bell14}). For a massless field the boundary-free part (\ref{FCsren})
vanishes, $\langle \bar{\psi}\psi \rangle _{\mathrm{s}}=0$. For massive
fields, the condensate $\langle \bar{\psi}\psi \rangle _{\mathrm{s}}$ is
positive everywhere. It diverges on the string as $1/r^{2}$ and is
exponentially suppressed at large distances, $mr\gg 1$.

The contribution of the term with $F(2mz,2my)$ in (\ref{SerSum}) to the
fermionic condensate\ does not depend on $a$ whereas the contribution of the
last term in the right-hand side of (\ref{SerSum}) vanishes in the limit $%
a\rightarrow \infty $. From here it follows that the part
\begin{eqnarray}
\langle \bar{\psi}\psi \rangle _{\mathrm{b}}^{(1)} &=&\frac{2m^{3}}{\pi ^{2}}%
\bigg[\sideset{}{'}{\sum}_{l=0}^{p}(-1)^{l}c_{l}F(2mz,2mrs_{l})+\frac{2q}{%
\pi }\cos \left( \frac{q\pi }{2}\right)  \notag \\
&&\times \int_{0}^{\infty }dx\frac{\cos \left( q\pi /2\right) \sinh \left(
qx\right) \sinh x}{\cosh (2qx)-\cos (q\pi )}F(2mz,2mr\cosh x)\bigg],
\label{FCB4}
\end{eqnarray}%
is induced by the presence of the boundary at $z=0$ when the second boundary
is absent. Hence, in the geometry of a single boundary at $z=0$ the
renormalized fermionic condensate is decomposed as%
\begin{equation}
\langle \bar{\psi}\psi \rangle ^{(1)}=\langle \bar{\psi}\psi \rangle _{%
\mathrm{s}}+\langle \bar{\psi}\psi \rangle _{\mathrm{b}}^{(1)}.
\label{FCdec}
\end{equation}%
This result, with (\ref{FCsren}) and (\ref{FCB4}), coincides with that
obtained in \cite{Beze13}. Hence, in the region between two boundaries the
fermionic condensate is written in the form%
\begin{eqnarray}
\langle \bar{\psi}\psi \rangle &=&\langle \bar{\psi}\psi \rangle ^{(1)}+%
\frac{1}{2\pi ^{2}}\left[ \sideset{}{'}{\sum}%
_{l=0}^{p}(-1)^{l}c_{l}B(z,rs_{l})\right.  \notag \\
&&\left. +\frac{2q}{\pi }\cos \left( \frac{q\pi }{2}\right) \int_{0}^{\infty
}dx\frac{\sinh \left( qx\right) \sinh (x)B(z,r\cosh x)}{\cosh (2qx)-\cos
(q\pi )}\right] ,  \label{FC3}
\end{eqnarray}%
where the last term is induced by the presence of the second boundary at $%
z=a $.

By taking into account that
\begin{equation}
F(2mz,2my)=-\frac{m^{-3}}{2y}\int_{m}^{\infty }dx\,\frac{x-m}{e^{2zx}}\sin
(2y\sqrt{x^{2}-m^{2}}),  \label{Frel}
\end{equation}%
and combining the single plate-induced part $\langle \bar{\psi}\psi \rangle
_{\mathrm{b}}^{(1)}$ with the last term in (\ref{FC3}), we find the
decomposition%
\begin{equation}
\langle \bar{\psi}\psi \rangle =\langle \bar{\psi}\psi \rangle _{\mathrm{s}%
}+\langle \bar{\psi}\psi \rangle _{\mathrm{b}}.  \label{FCdec2}
\end{equation}%
Here, the boundary-induced contribution to the fermionic condensate in the
region between the plates is given by
\begin{eqnarray}
\langle \bar{\psi}\psi \rangle _{\mathrm{b}} &=&\frac{1}{2\pi ^{2}}\left[ %
\sideset{}{'}{\sum}_{l=0}^{p}(-1)^{l}c_{l}C(z,rs_{l})+\frac{2q}{\pi }\cos
\left( \frac{q\pi }{2}\right) \right.   \notag \\
&&\left. \times \int_{0}^{\infty }dx\frac{\sinh \left( qx\right) \sinh
(x)C(z,r\cosh x)}{\cosh (2qx)-\cos (q\pi )}\right] ,  \label{FCC}
\end{eqnarray}%
with the function%
\begin{equation}
C(z,y)=\frac{2}{y}\int_{m}^{\infty }dx\frac{\sin (2y\sqrt{x^{2}-m^{2}})}{%
\frac{x+m}{x-m}e^{2ax}+1}\left[ 2m-(m+x)\left( e^{2zx}+e^{2(a-z)x}\right) %
\right] .  \label{Czy}
\end{equation}%
Taking the limit $a\rightarrow \infty $, $z\rightarrow \infty $, with fixed $%
a-z$, and using the relation (\ref{Frel}), we can see that from the second
term in the right-hand side of (\ref{FC3}) the boundary-induced part is
obtained for a single plate at $z=a$. The latter is given by the expression (%
\ref{FCB4}) with the replacement $z\rightarrow a-z$.

For a massless field, $m=0$, from (\ref{Czy}) one gets%
\begin{equation}
C(z,y)=\frac{\pi }{ay}\partial _{z}\frac{\sinh (\pi y/a)\cos (\pi z/a)}{%
\cosh (2\pi y/a)-\cos (2\pi z/a)}.  \label{Czym0}
\end{equation}%
In this special case the boundary-free part vanishes and for the single
plate contribution one has (see \cite{Beze13})%
\begin{equation}
\langle \bar{\psi}\psi \rangle ^{(1)}=-\frac{q}{4\pi ^{2}rz^{2}}\frac{u^{q}}{%
u^{2q}-1}\left( 1+\frac{qz}{\sqrt{r^{2}+z^{2}}}\frac{u^{2q}+1}{u^{2q}-1}%
\right) ,  \label{FCbm0}
\end{equation}%
with the notation $u=z/r+\sqrt{1+z^{2}/r^{2}}$.

For $q=1$, in (\ref{FCC}) the $l=0$ term survives only and we get the
corresponding result for two parallel plates in Minkowski bulk:%
\begin{equation}
\langle \bar{\psi}\psi \rangle _{\mathrm{M}}=\frac{1}{\pi ^{2}}%
\int_{m}^{\infty }dx\frac{\sqrt{x^{2}-m^{2}}}{\frac{x+m}{x-m}e^{2ax}+1}\left[
2m-(m+x)\left( e^{2zx}+e^{2(a-z)x}\right) \right] .  \label{FCMink}
\end{equation}%
In this case the fermionic condensate is negative. The remaining part in (%
\ref{FCC}) is induced by the presence of the cosmic string. In particular,
for a massless field the Minkowskian part is simplified to%
\begin{equation}
\langle \bar{\psi}\psi \rangle _{\mathrm{M}}=-\pi \frac{2-\sin ^{2}(\pi z/a)%
}{8a^{3}\sin ^{3}(\pi z/a)}.  \label{FCMinkm0}
\end{equation}%
For points near the plate $z=0$, to the leading order, we get%
\begin{equation}
\langle \bar{\psi}\psi \rangle _{\mathrm{M}}\approx -\frac{1}{4\pi ^{2}z^{3}}%
.  \label{FCMz0}
\end{equation}%
This expression gives the leading term in the asymptotic expansion for the
fermionic condensate near the plate for massive fields as well. Moreover, to
the leading order and under the condition $z\ll r$, the expression in the
right-hand side of (\ref{FCMz0}) gives the asymptotic in the presence of the
cosmic string: $\langle \bar{\psi}\psi \rangle \approx -1/(4\pi ^{2}z^{3})$.
The generalization of (\ref{FCMink}) for arbitrary number of spatial
dimensions is given in \cite{Eliz11}. The fermionic condensate for the
geometry of two parallel on AdS bulk has been investigated in \cite{Eliz13}.

For points outside the plates, $z\neq 0,a$, the boundary-induced
contribution (\ref{FCC}) is finite on the string. By taking into account
that $\lim_{y\rightarrow 0}C(z,y)=4\pi ^{2}\langle \bar{\psi}\psi \rangle
_{M}$, we obtain%
\begin{equation}
\langle \bar{\psi}\psi \rangle _{\mathrm{b},r=0}=2\langle \bar{\psi}\psi
\rangle _{M}\left[ \sideset{}{'}{\sum}_{l=0}^{p}(-1)^{l}c_{l}+\frac{2q}{\pi }%
\cos \left( \frac{q\pi }{2}\right) \int_{0}^{\infty }dx\frac{\sinh \left(
qx\right) \sinh x}{\cosh (2qx)-\cos (q\pi )}\right] .  \label{FCbr0}
\end{equation}%
The expression in the square brackets vanishes and, hence, the
boundary-induced part in the fermionic condensate vanishes on the string.
From here it follows that for points near the string and for a massive field
the total fermionic condensate is dominated by the boundary-free
contribution and is positive.

Now let us consider the asymptotic behavior of the fermionic condensate at
large distances from the string, $r\gg a,m^{-1}$. Here we need the
asymptotic of the function $C(z,y)$ for large values of the second argument.
It is obtained by taking into account that for large $y$ the dominant
contribution to the integral in (\ref{Czy}) comes from the region near the
lower limit of the integration. To the leading order, the integral is
expressed in terms of the Macdonald function. By using the corresponding
asymptotic expression for large arguments, we can see that%
\begin{equation}
C(z,y)\approx \sqrt{\pi }m^{3}\left[ e^{2zm}+e^{2(a-z)m}-1\right] \frac{%
e^{-2my}}{(my)^{3/2}},  \label{CzyLarge}
\end{equation}%
for $y\gg a,m^{-1}$. At large distances, the leading contribution in the
boundary-induced part of the fermionic condensate comes from the term $l=0$
which coincides with $\langle \bar{\psi}\psi \rangle _{\mathrm{M}}$. For $q>2
$, the next to the leading term comes from the term $l=1$ in (\ref{FCC}).
Adding the asymptotic for the boundary-free contribution one gets%
\begin{equation}
\langle \bar{\psi}\psi \rangle \approx \langle \bar{\psi}\psi \rangle _{%
\mathrm{M}}-\left[ e^{2zm}+e^{2(a-z)m}-2\right] \frac{m^{3}\cos \left( \pi
/q\right) e^{-2mr\sin (\pi /q)}}{2(\pi mr\sin (\pi /q))^{3/2}}.
\label{FClarge}
\end{equation}%
The topological part, given by the second term in the right-hand side, is
negative. For a massless field the asymptotic expression is directly
obtained by using (\ref{Czym0}):%
\begin{equation}
\langle \bar{\psi}\psi \rangle \approx \langle \bar{\psi}\psi \rangle _{%
\mathrm{M}}+\frac{\cot (\pi /q)\sin (\pi z/a)}{2\pi a^{2}re^{\pi r\sin (\pi
/q)/a}}.  \label{FClargem0}
\end{equation}%
Note that in this case the topological part is positive. For $q\leqslant 2$,
the decay of the topological part in the boundary-induced fermion condensate
at large distances from the string is stronger, like $e^{-2mr}$ and $e^{-\pi
r/a}$ for the massive and massless cases, respectively. Hence, at large
distances from the string, to the leading order, the fermionic condensate
coincides with that for the plates in Minkowski bulk and is negative. As we
have shown before, the condensate is positive near the string. As a
consequence, the fermionic condensate vanishes for some intermediate value
of the radial coordinate $r$. Note that at large distances from the string
the topological part in the region between the plates is suppressed
exponentially for both massive and massless fields.

\begin{figure}[tbph]
\begin{center}
\epsfig{figure=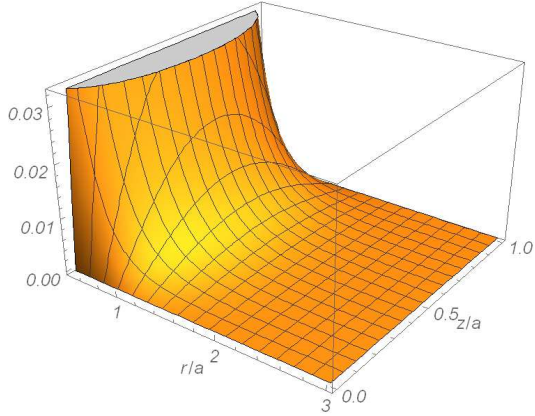,width=7.cm,height=6.cm}
\end{center}
\caption{The topological part in the fermionic condensate, $a^{3}\langle
\bar{\protect\psi}\protect\psi \rangle _{\mathrm{cs}}$, for a massless field
as a function of the distances from the string and from the left plate. For
the planar angle deficit we have taken the value corresponding to $q=3$.}
\label{fig1}
\end{figure}

Let us denote the part in the fermionic condensate induced by the cosmic
string as $\langle \bar{\psi}\psi \rangle _{\mathrm{t}}$. It is given by $%
\langle \bar{\psi}\psi \rangle _{\mathrm{t}}=\langle \bar{\psi}\psi \rangle
-\langle \bar{\psi}\psi \rangle _{\mathrm{M}}$ and can be termed as
topological part. The corresponding expression is obtained from (\ref{FCdec2}%
) omitting the $l=0$ term in the boundary-induced contribution (\ref{FCC}).
For a massless field, by taking into account that for $y\neq 0$ one has $%
C(0,y)=C(a,y)=0$ (see (\ref{Czym0})), we conclude that the topological part
in the fermionic condensate vanishes on the plates. The same is the case for
a massive field. In order to show that, we firstly consider the topological
part in the boundary-induced contribution (\ref{FCC}). This part is obtained
from (\ref{FCC}) omitting the term $l=0$. For the evaluation of the
topological part we cannot directly put $z=0,a$ in the corresponding
expression. The limit $z\rightarrow 0,a$ should be taken after the
evaluation of the integral for $z\neq 0,a$. In fact, we need to evaluate the
limits $\lim_{z\rightarrow 0,a}C(z,y)$ for $y\neq 0$. These limits are the
same and we will consider the case $z\rightarrow 0$ only. The evaluation
procedure is simplified if we introduce in the integrand the factor $%
e^{-\alpha \sqrt{x^{2}-m^{2}}}$ with $\alpha >0$ and take the limit $\alpha
\rightarrow 0$ after the evaluation of the integral. With this factor we can
directly put $z=0$ in the integrand. In this way it can be seen that%
\begin{equation}
\lim_{z\rightarrow 0}C(z,y)=4m^{3}f_{1}(2my),  \label{limC}
\end{equation}%
for $y\neq 0$. Comparing with the boundary-free fermionic condensate (\ref%
{FCsren}), we see that the boundary-induced contribution in the topological
part exactly cancels the boundary-free part and the topological part of the
fermionic condensate vanishes on the plates. This feature is seen from
figure \ref{fig1}, where, for a massless field and for the value of the
parameter $q=3$, we have plotted the topological part in the fermionic
condensate, $a^{3}\langle \bar{\psi}\psi \rangle _{\mathrm{t}}$, as a
function of the distance from the string and of the distance from the plate
at $z=0$.

An alternative representation for the fermionic condensate is obtained by
the application of the summation formula (\ref{Sumxn}) to the series over $n$
in (\ref{FCin}). The part with the first integral in the right-hand side of (%
\ref{Sumxn}) will give the fermion condensate for the geometry of a single
plate at $z=0$. In the corresponding expression we first separate the
boundary-free contribution, obtained from the first term in the right-hand
side of\ (\ref{hkz}), and in the remaining part we rotate the integration
contour over $k$ by the angle $\pi /2$ ($-\pi /2$) for the term with $%
e^{2ikz}$ ($e^{-2ikz}$). Combining with the term coming from the second
integral in (\ref{Sumxn}), for the fermionic condensate in the region
between the plates we obtain the representation%
\begin{eqnarray}
\langle \bar{\psi}\psi \rangle &=&\langle \bar{\psi}\psi \rangle _{\mathrm{s}%
}-\frac{q}{\pi ^{2}}\sum_{j}\int_{0}^{\infty }d\lambda \,\lambda \left[
J_{qj-1/2}^{2}(\lambda r)+J_{qj+1/2}^{2}(\lambda r)\right]  \notag \\
&&\times \int_{\sqrt{\lambda ^{2}+m^{2}}}^{\infty }dx\frac{\left( x+m\right) %
\left[ e^{2xz}+e^{2(a-z)x}\right] -2m}{\sqrt{x^{2}-\lambda ^{2}-m^{2}}\left(
\frac{x+m}{x-m}e^{2ax}+1\right) }.  \label{FCalt}
\end{eqnarray}%
As is seen from this formula, the boundary-induced contribution to the
fermionic condensate, given by the second term in the right-hand side of (%
\ref{FCalt}), is always negative. In the limit $r\rightarrow 0$, the
dominant contribution to the boundary-induced part in (\ref{FCalt}) comes
from the term $j=1/2$ and, by using the asymptotic for the Bessel function
for small values of the argument, to the leading order we get%
\begin{eqnarray}
\langle \bar{\psi}\psi \rangle &\approx &\langle \bar{\psi}\psi \rangle _{%
\mathrm{s}}-\frac{\pi ^{-3/2}(r/2)^{q-1}}{\Gamma (q/2)\Gamma ((q+1)/2)}%
\int_{m}^{\infty }dx\,\frac{(x^{2}-m^{2})^{q/2}}{\frac{x+m}{x-m}e^{2ax}+1}
\notag \\
&&\times \left[ \left( x+m\right) (e^{2xz}+e^{2(a-z)x})-2m\right] .
\label{FCr0b}
\end{eqnarray}%
For $q=1$ this asymptotic formula is reduced to the exact result (\ref%
{FCMink}).

In a number of field theoretical models, the Lagrangian, in addition to the
fermionic mass term $m\bar{\psi}\psi $, contains other terms involving the
product $\bar{\psi}\psi $. An example is the fermionc field nonminimally
coupled to gravity with the interaction term $\xi R\bar{\psi}\psi $ (see,
e.g., \cite{Riba08}), where $R$ is the scalar curvature for the background
spacetime and $\xi $ is a constant with the dimension of the inverse mass.
The formation of nonzero fermionic condensate induces an additional term $%
\xi R\langle \bar{\psi}\psi \rangle $ in the Lagrangian for the
gravitational field. This leads to the shift (in general, space-dependent)
in the gravitational constant. Another example is the fermionic field
interacting with a scalar field $\varphi $ through the interaction
Lagrangian proportional to $\varphi \bar{\psi}\psi $. The interaction terms
of this form appear also in semibosonized versions of the four-fermion
interaction models of the Gross-Neveu and Nambu-Jona-Lasinio type. The
effects of the background geometry, topology and boundaries in these models
have been discussed in \cite{Inag97,Eber10}.

\section{Energy-momentum tensor}

\label{sec:EMT}

The mode sum for the VEV\ of the energy-momentum tensor, $\left\langle
0\left\vert T_{\mu \nu }\right\vert 0\right\rangle \equiv \left\langle
T_{\mu \nu }\right\rangle $, is given by the expression
\begin{equation}
\left\langle T_{\mu \nu }\right\rangle =\frac{i}{2}\sum_{s=\pm 1}\sum_{j=\pm
1/2,\ldots }\sum_{n=1}^{\infty }\int_{0}^{\infty }d\lambda \,[\bar{\psi}%
_{\sigma }^{(-)}\gamma _{(\mu }\nabla _{\nu )}\psi _{\sigma }^{(-)}-(\nabla
_{(\mu }\bar{\psi}_{\sigma }^{(-)})\gamma _{\nu )}\psi _{\sigma }^{(-)}]\ ,
\label{modesum}
\end{equation}%
where $\gamma _{\mu }=g_{\mu \nu }\gamma ^{\nu }$ and the brackets enclosing
the indices mean the symmetrization. Similar to the case of the condensate,
we assume that some regularization procedure is used for the right-hand side
of (\ref{modesum}) without explicitly writing it. Inserting the expression
for the negative-energy mode functions, we can see that the off-diagonal
components vanish. The VEVs for the diagonal components are presented in the
form (no summation over $\nu $):%
\begin{equation}
\left\langle T_{\nu }^{\nu }\right\rangle =\frac{q}{\pi a}%
\sum_{j}\sum_{n=1}^{\infty }\int_{0}^{\infty }d\lambda \,\frac{\lambda ^{3}}{%
\omega }\frac{g_{qj-1/2}^{(\nu )}(\lambda r)h_{\nu }(k_{n},z)}{1-\sin
(2x_{n})/(2x_{n})},  \label{Tnu}
\end{equation}%
with the notations%
\begin{eqnarray}
g_{\beta }^{(0)}(y) &=&-\frac{\omega ^{2}}{k^{2}}g_{\beta }^{(3)}(y)=-\frac{%
\omega ^{2}}{\lambda ^{2}}[J_{\beta }^{2}(y)+J_{\beta +1}^{2}(y)],  \notag \\
g_{\beta }^{(1)}(y) &=&J_{\beta }^{2}(y)+J_{\beta +1}^{2}(y)-\frac{2qj}{y}%
J_{\beta }(y)J_{\beta +1}(y),  \label{gebet} \\
g_{\beta }^{(2)}(y) &=&\frac{2qj}{y}J_{\beta }(y)J_{\beta +1}(y),  \notag
\end{eqnarray}%
and%
\begin{equation}
h_{\nu }(k,z)=1-\frac{1-\delta _{3}^{\nu }}{2}\sum_{\eta =\pm 1}\frac{%
me^{2\eta ikz}}{m-\eta ik}.  \label{hnu}
\end{equation}%
Note that the axial stress does not depend on the coordinate $z$. We can see
that the VEVs (\ref{Tnu}) obey the trace relation
\begin{equation}
\left\langle T_{\nu }^{\nu }\right\rangle =m\langle \bar{\psi}\psi \rangle .
\label{TrRel}
\end{equation}

In Appendix we show that the diagonal components of the vacuum
energy-momentum tensor are transformed to the form%
\begin{eqnarray}
\left\langle T_{\nu }^{\nu }\right\rangle  &=&-\frac{1}{4\pi a}%
\sum_{n=1}^{\infty }\frac{1}{1-\sin (2x_{n})/(2x_{n})}\bigg[%
\sideset{}{'}{\sum}_{l=0}^{p}(-1)^{l}c_{l}D_{\nu }(z,rs_{l},k_{n})  \notag \\
&&+\frac{2q}{\pi }\cos \left( \frac{q\pi }{2}\right) \int_{0}^{\infty }du\,%
\frac{\sinh \left( qu\right) \sinh (u)D_{\nu }(z,r\cosh u,k_{n})}{\cosh
(2qu)-\cos (q\pi )}\bigg],  \label{TnuComb}
\end{eqnarray}%
with the notation%
\begin{equation*}
D_{\nu }(z,y,k)=\hat{D}_{\nu }e^{-2y\sqrt{k^{2}+m^{2}}}h_{\nu }(k,z),
\end{equation*}%
where we have defined the operators%
\begin{eqnarray}
\hat{D}_{0} &=&\hat{D}_{1}=y^{-3}(y\partial _{y}-1),  \notag \\
\hat{D}_{2} &=&2y^{-3}(1-y\partial _{y}+\frac{1}{2}y^{2}\partial _{y}^{2}),
\notag \\
\hat{D}_{3} &=&y^{-1}\left( 4m^{2}-\partial _{y}^{2}\right) .  \label{Dnu}
\end{eqnarray}%
For these operators we have the relation%
\begin{equation}
\sum_{\nu =0}^{3}\hat{D}_{\nu }=4m^{2}/y.  \label{DnuSum}
\end{equation}%
As is seen from (\ref{Dnu}), the radial stress is equal to the energy
density.

Now we apply to the series over $n$ in (\ref{TnuComb}) the summation formula
(\ref{Sumxn}) with the function
\begin{equation}
f(x)=e^{-2y\sqrt{x^{2}/a^{2}+m^{2}}}h_{\nu }(x/a,z).  \label{fhnu}
\end{equation}%
As a result, the VEVs are decomposed as%
\begin{eqnarray}
\left\langle T_{\nu }^{\nu }\right\rangle &=&\left\langle T_{\nu }^{\nu
}\right\rangle ^{(1)}+\frac{1}{2\pi ^{2}}\bigg[\sideset{}{'}{\sum}%
_{l=0}^{p}(-1)^{l}c_{l}B_{\nu }(z,rs_{l})  \notag \\
&&+\frac{2q}{\pi }\cos \left( \frac{q\pi }{2}\right) \int_{0}^{\infty }du\,%
\frac{\sinh \left( qu\right) \sinh (u)B_{\nu }(z,r\cosh u)}{\cosh (2qu)-\cos
(q\pi )}\bigg],  \label{TnuDec}
\end{eqnarray}%
with the function%
\begin{equation}
B_{\nu }(z,y)=\hat{D}_{\nu }\int_{m}^{\infty }dx\frac{\sin (2y\sqrt{%
x^{2}-m^{2}})}{\frac{x+m}{x-m}e^{2ax}+1}\left( 1-\frac{1-\delta _{3}^{\nu }}{%
2}\sum_{\eta =\pm 1}\frac{me^{2\eta xz}}{m-\eta x}\right) .  \label{Benu}
\end{equation}%
The first term in the right-hand side of (\ref{TnuDec}) comes from the first
integral in (\ref{Sumxn}) and is given by the expression%
\begin{eqnarray}
\left\langle T_{\nu }^{\nu }\right\rangle ^{(1)} &=&-\frac{1}{4\pi ^{2}}%
\bigg[\sideset{}{'}{\sum}_{l=0}^{p}(-1)^{l}c_{l}B_{\nu }^{(0)}(z,rs_{l})+%
\frac{2q}{\pi }\cos \left( \frac{q\pi }{2}\right)  \notag \\
&&\times \int_{0}^{\infty }du\,\frac{\sinh \left( qu\right) \sinh (u)B_{\nu
}^{(0)}(z,r\cosh u)}{\cosh (2qu)-\cos (q\pi )}\bigg],  \label{T1nu}
\end{eqnarray}%
where%
\begin{equation}
B_{\nu }^{(0)}(z,y)=\hat{D}_{\nu }\int_{0}^{\infty }dx\,e^{-2y\sqrt{%
x^{2}+m^{2}}}h_{\nu }(x,z).  \label{Bnu0}
\end{equation}%
The contribution (\ref{T1nu}) does not depend on $a$, whereas the second
term in the right-hand side of (\ref{TnuDec}) vanishes in the limit $%
a\rightarrow \infty $. From here it follows that the part $\left\langle
T_{\nu }^{\nu }\right\rangle ^{(1)}$ corresponds to the VEV in the geometry
of a single plate at $z=0$.

The further transformation of the contribution $\left\langle T_{\nu }^{\nu
}\right\rangle ^{(1)}$ is similar to that we have used for $\langle \bar{\psi%
}\psi \rangle ^{(1)}$ in the previous section. We decompose $\left\langle
T_{\nu }^{\nu }\right\rangle ^{(1)}$ into two parts coming from the first
and second terms in the right-hand side of (\ref{hnu}). The part with $%
h_{\nu }(x,z)\rightarrow 1$ corresponds to the VEV in the boundary-free
cosmic string spacetime. In this part the corresponding integral (\ref{Bnu0}%
) is expressed in terms of the Macdonald function and the renormalization is
reduced to omitting the $l=0$ term. The corresponding renormalizaed VEV is
presented as (no summation over $\nu $)%
\begin{eqnarray}
\left\langle T_{\nu }^{\nu }\right\rangle _{\mathrm{s}} &=&\frac{2m^{4}}{\pi
^{2}}\bigg[\sum_{l=1}^{p}(-1)^{l}c_{l}F_{\nu }^{(0)}(2mrs_{l})+\frac{2q}{\pi
}\cos \left( \frac{q\pi }{2}\right)   \notag \\
&&\times \int_{0}^{\infty }dx\frac{\sinh \left( qx\right) \sinh (x)F_{\nu
}^{(0)}(2mr\cosh x)}{\cosh (2qx)-\cos (q\pi )}\bigg],  \label{EMTsren1}
\end{eqnarray}%
where we have defined the functions%
\begin{eqnarray}
F_{0}^{(0)}(y) &=&F_{1}^{(0)}(y)=F_{3}^{(0)}(y)=f_{2}(y),  \notag \\
F_{2}^{(0)}(y) &=&-3f_{2}(y)-f_{1}(y),  \label{F20}
\end{eqnarray}%
with the notation (\ref{fnu}). The corresponding energy density is negative
everywhere. An alternative integral representation is given in \cite{Beze08}%
. The fermionic VEVs $\left\langle T_{\nu }^{\nu }\right\rangle _{\mathrm{s}}
$ in the special case $q<2$ have been previously investigated in \cite%
{Beze06}. For a massless field the renormalized VEV of the energy-momentum
tensor in the boundary-free cosmic string spacetime was found in \cite%
{Frol87} and is obtained from (\ref{EMTsren1}) taking the limit $%
m\rightarrow 0$:%
\begin{equation}
\left\langle T_{\nu }^{\nu }\right\rangle _{\mathrm{s}}=-\frac{%
(q^{2}-1)(7q^{2}+17)}{2880\pi ^{2}r^{4}}\mathrm{diag}(1,1,-3,1).
\label{Tnusren}
\end{equation}%
Note that in the boundary-free cosmic string spacetime one has the relation $%
\left\langle T_{0}^{0}\right\rangle _{\mathrm{s}}=\left\langle
T_{3}^{3}\right\rangle _{\mathrm{s}}$. The latter is a consequence of the
boost invariance along the axis of the cosmic string. For a massive field
and for points near the string, $mr\ll 1$, the renormalized VEV diverges on
the string as $r^{-4}$ with the leading term given by (\ref{Tnusren}). At
large distances, $mr\gg 1$, for $q<2$ the first term in the square brackets
in (\ref{EMTsren1}) is absent and the dominant contribution in the second
term comes from the region near the lower limit of the integration. In this
case $\left\langle T_{\nu }^{\nu }\right\rangle _{\mathrm{s}}$ is suppressed
by the factor $e^{-2mr}$. For $q>2$ the dominant contribution at large
distances comes from the $l=1$ term in (\ref{EMTsren1}) and the suppression
factor is $e^{-2mr\sin (\pi /q)}$. Note that the contribution of the first
term in the square brackets of (\ref{EMTsren1}) is always negative for $\nu
=0,1,3$, and positive for $\nu =2$. It can be easily checked that the
boundary-free contribution (\ref{EMTsren1}) separately obeys the trace
relation (\ref{TrRel}).

The contribution in (\ref{T1nu}), coming from the second term in the
right-hand side of (\ref{hnu}), is induced by the boundary at $z=0$ when the
second boundary is absent. After the rotation of the integration contours in
the integral (\ref{Bnu0}) one gets%
\begin{equation}
\int_{0}^{\infty }dx\,e^{-2y\sqrt{x^{2}+m^{2}}}\sum_{\eta =\pm 1}\frac{%
me^{2\eta ikz}}{m-\eta ik}=2mG(2mz,2my),  \label{RelInt}
\end{equation}%
where%
\begin{equation}
G(2mz,2my)=\int_{m}^{\infty }dx\,\frac{e^{-2xz}}{m+x}\sin (2y\sqrt{%
x^{2}-m^{2}}).  \label{Cm}
\end{equation}%
An equivalent expression is obtained by using the representation $%
(m+x)^{-1}=\int_{0}^{\infty }dt\,e^{-(m+x)t}$. The integral over $x$ is
expressed in terms of the Macdonald function and we find%
\begin{equation}
G(u,v)=vG_{1}(u,v).  \label{Cu}
\end{equation}%
Here and below we use the notation%
\begin{equation}
G_{\nu }(u,v)=e^{u}\int_{u}^{\infty }dx\,e^{-x}f_{\nu }(\sqrt{x^{2}+v^{2}}).
\label{Gnu}
\end{equation}%
As a result, the VEV in the geometry of the single plate at $z=0$ is
decomposed as (no summation over $\nu $)%
\begin{equation}
\left\langle T_{\nu }^{\nu }\right\rangle ^{(1)}=\left\langle T_{\nu }^{\nu
}\right\rangle _{\mathrm{s}}+\left\langle T_{\nu }^{\nu }\right\rangle _{%
\mathrm{b}}^{(1)},  \label{EMTdec}
\end{equation}%
where the boundary-induced contribution is given by
\begin{eqnarray}
\left\langle T_{\nu }^{\nu }\right\rangle _{\mathrm{b}}^{(1)} &=&\frac{2m^{4}%
}{\pi ^{2}}\bigg[\sideset{}{'}{\sum}_{l=0}^{p}(-1)^{l}c_{l}F_{\nu
}(2mz,2mrs_{l})+\frac{2q}{\pi }\cos \left( \frac{q\pi }{2}\right)  \notag \\
&&\times \int_{0}^{\infty }dx\frac{\sinh \left( qx\right) \sinh x}{\cosh
(2qx)-\cos (q\pi )}F_{\nu }(2mz,2mr\cosh x)\bigg],  \label{EMTb1}
\end{eqnarray}%
with the functions%
\begin{eqnarray}
F_{0}(x,y) &=&F_{1}(x,y)=-G_{2}(x,y),  \notag \\
F_{2}(x,y) &=&2G_{2}(x,y)+F(x,y),  \label{F01n}
\end{eqnarray}%
and $F_{3}(x,y)=0$. As we see, the boundary-induced part in the axial stress
vanishes. Note that we have the relation%
\begin{equation}
F_{\nu }(2mz,2my)=\frac{1-\delta _{3}^{\nu }}{8m^{3}}\hat{D}_{\nu
}G(2mz,2my).  \label{FnuRel}
\end{equation}%
By using this relation, we can see that in the limit $a\rightarrow \infty $,
$z\rightarrow \infty $, with fixed $a-z$, from the second term in the
right-hand side of (\ref{TnuDec}) we obtain the boundary-induced part in the
geometry of a single plate at $z=a$. The latter is given by (\ref{EMTb1})
with the replacement $z\rightarrow a-z$.

By taking into account (\ref{FnuRel}) and combining the last term in the
right-hand side of (\ref{TnuDec}) with the single plate-induced part (\ref%
{EMTb1}), we get the following representation%
\begin{equation}
\left\langle T_{\nu }^{\nu }\right\rangle =\left\langle T_{\nu }^{\nu
}\right\rangle _{\mathrm{s}}+\left\langle T_{\nu }^{\nu }\right\rangle _{%
\mathrm{b}}.  \label{TnuDec2p}
\end{equation}%
Here the boundary-free part is given by (\ref{EMTsren1}) and for the
boundary-induced contribution in the region between the plates one has the
expression%
\begin{eqnarray}
\left\langle T_{\nu }^{\nu }\right\rangle _{\mathrm{b}} &=&\frac{1}{2\pi ^{2}%
}\bigg[\sideset{}{'}{\sum}_{l=0}^{p}(-1)^{l}c_{l}C_{\nu }(z,rs_{l})+\frac{2q%
}{\pi }\cos \left( \frac{q\pi }{2}\right)  \notag \\
&&\times \int_{0}^{\infty }du\,\frac{\sinh \left( qu\right) \sinh (u)C_{\nu
}(z,r\cosh u)}{\cosh (2qu)-\cos (q\pi )}\bigg],  \label{TnuDec2}
\end{eqnarray}%
with the notation%
\begin{equation}
C_{\nu }(z,y)=\hat{D}_{\nu }\int_{m}^{\infty }dx\frac{\sin (2y\sqrt{%
x^{2}-m^{2}})}{\frac{x+m}{x-m}e^{2ax}+1}\left[ 1+\,\frac{m}{2}\left(
1-\delta _{3}^{\nu }\right) \frac{e^{2xz}+e^{2(a-z)x}}{x-m}\right] .
\label{Cnu}
\end{equation}%
As we could expect, in the region between the plates the VEVs are symmetric
with respect to the plane $z=a/2$. For a massless field one obtains%
\begin{equation}
C_{\nu }(z,y)=\frac{\pi }{4a}\hat{D}_{\nu }\left[ \frac{1}{\pi y/a}-\frac{1}{%
\sinh (\pi y/a)}\right] .  \label{Cnum0}
\end{equation}%
Note that in the latter case the single plate part in the VEV of the
energy-momentum tensor vanishes and in the region between the plates the VEV
of the energy-momentum tensor does not depend on the coordinate $z$.

It can be checked that the boundary-induced contribution obeys the covariant
conservation equation $\nabla _{\mu }\left\langle T_{\nu }^{\mu
}\right\rangle _{\mathrm{b}}=0$. For the geometry under consideration the
latter is reduced to a single equation $\partial _{r}\left( r\left\langle
T_{1}^{1}\right\rangle _{\mathrm{b}}\right) =\left\langle
T_{2}^{2}\right\rangle _{\mathrm{b}}$. The latter equation directly follows
from the relation%
\begin{equation}
\left( y\partial _{y}+1\right) \hat{D}_{1}\sin (by)=\hat{D}_{2}\sin (by).
\label{RelCont}
\end{equation}%
By taking into account the relation (\ref{DnuSum}) we can see that $%
\sum_{\nu =0}^{3}C_{\nu }(z,y)=mC(z,y)$. From here it follows that the
boundary induced contributions obey the trace relation (\ref{TrRel}).

For the further transformation of the VEV\ (\ref{TnuDec2}) in the case of a
massless field, we first separate the Minkowskian part (the $l=0$ term). It
can be seen that the contribution of the first term in the square brackets
of (\ref{Cnum0}) to the remaining (topological) part of $\left\langle T_{\nu
}^{\nu }\right\rangle _{\mathrm{b}}$ is equal to $-\left\langle T_{\nu
}^{\nu }\right\rangle _{\mathrm{s}}$. Hence, in the total VEV, this
contribution is cancelled by the boundary-free part and for the total VEV in
the massless case we get%
\begin{eqnarray}
\left\langle T_{\nu }^{\nu }\right\rangle  &=&\left\langle T_{\nu }^{\nu
}\right\rangle _{\mathrm{M}}-\frac{\pi ^{2}}{8a^{4}}\bigg[%
\sum_{l=1}^{p}(-1)^{l}c_{l}A_{\nu }(\pi s_{l}r/a)+\frac{2q}{\pi }\cos \left(
\frac{q\pi }{2}\right)   \notag \\
&&\times \int_{0}^{\infty }du\,\frac{\sinh \left( qu\right) \sinh (u)A_{\nu
}(\pi (r/a)\cosh u)}{\cosh (2qu)-\cos (q\pi )}\bigg],  \label{Tnum0}
\end{eqnarray}%
with the functions%
\begin{equation}
A_{\nu }(y)=\hat{D}_{\nu }\frac{1}{\sinh y}.  \label{Anu}
\end{equation}%
For a massless field from (\ref{DnuSum}) one has $\sum_{\nu =0}^{3}\hat{D}%
_{\nu }=0$ and both the Minkowskian and topological parts in (\ref{Tnum0})
are traceless.

In the absence of the cosmic string, the only nonzero contribution to (\ref%
{TnuDec2}) comes from the term with $l=0$ and we obtain the VEVs in the
region between two plates in Minkowski bulk (no summation over $\nu $):%
\begin{eqnarray}
\left\langle T_{\nu }^{\nu }\right\rangle _{\mathrm{M}} &=&-\frac{2}{3\pi
^{2}}\int_{m}^{\infty }dx\frac{(x^{2}-m^{2})^{3/2}}{\frac{x+m}{x-m}e^{2ax}+1}%
\left[ 1+\,\frac{m}{2}\frac{e^{2xz}+e^{2(a-z)x}}{x-m}\right] ,  \notag \\
\left\langle T_{3}^{3}\right\rangle _{\mathrm{M}} &=&\frac{2}{\pi ^{2}}%
\int_{m}^{\infty }dx\frac{x^{2}\sqrt{x^{2}-m^{2}}}{\frac{x+m}{x-m}e^{2ax}+1},
\label{T33Mink}
\end{eqnarray}%
where $\nu =0,1,2$. In particular, the energy density is negative. For a
massless field one has $\left\langle T_{3}^{3}\right\rangle _{\mathrm{M}%
}=-3\left\langle T_{0}^{0}\right\rangle _{\mathrm{M}}=7\pi ^{2}/(960a^{4})$.
The fermionic Casimir densities for the geometry of two parallel plates in
background of flat spacetime with an arbitrary number toroidally
compactified spatial dimensions have been investigated in \cite{Eliz11}. The
expressions (\ref{T33Mink}) are special cases of the corresponding general
formulas. The VEV\ of the energy-momentum tensor for parallel branes on AdS
bulk is investigated in \cite{Eliz13}. The fermionic Casimir energy for two
parallel plates in 4-dimensional Minkowski spacetime has been studied in
\cite{John75,Mama80}. The corresponding result for an arbitrary number of
dimensions is generalized in \cite{Paol99}. The influence of the
compactification of spatial dimensions on the fermionic Casimir energy has
been discussed in \cite{Bell09}.

Let us consider the boundary-induced contribution in the VEV of the
energy-momentum tensor in the asymptotic regions. For points outside the
plates, the boundary-induced part is finite on the string. Taking the limit $%
r\rightarrow 0$ it can be seen that $C_{\nu }(z,0)=4\pi ^{2}\left\langle
T_{\nu }^{\nu }\right\rangle _{\mathrm{M}}$. Hence, from (\ref{TnuDec2}) one
gets%
\begin{equation}
\left\langle T_{\nu }^{\nu }\right\rangle _{\mathrm{b},r=0}=2\left\langle
T_{\nu }^{\nu }\right\rangle _{\mathrm{M}}\left[ \sideset{}{'}{\sum}%
_{l=0}^{p}(-1)^{l}c_{l}+\frac{2q}{\pi }\cos \left( \frac{q\pi }{2}\right)
\int_{0}^{\infty }dx\,\frac{\sinh \left( qx\right) \sinh x}{\cosh (2qx)-\cos
(q\pi )}\right] .  \label{Tbr0}
\end{equation}%
Note that the expression in the square brackets is the same as that in the
corresponding formula (\ref{FCbr0}) for the fermionic condensate. This
expression is zero and, hence, the boundary-induced contribution in the VEV
of the energy-momentum tensor vanishes on the string. This means that in the
region near the cosmic string the VEV of the energy-momentum tensor is
dominated by the boundary-free part and the corresponding energy density is
negative.

In order to obtain the behavior of the vacuum energy-momentum tensor at
large distances from the string, assuming that $mr\gg 1$, we need to find
the asymptotic estimate for the function $C_{\nu }(z,y)$ for large values of
$y$. In this limit the dominant contribution to the integral in (\ref{Cnu})
comes from the region near the lower limit of the integration and one gets%
\begin{eqnarray}
C_{\nu }(z,y) &\approx &\frac{e^{2mz}+e^{2m(a-z)}}{4m}\hat{D}_{\nu
}\int_{m}^{\infty }dx\,x\sin (2y\sqrt{x^{2}-m^{2}})e^{-2ax},  \notag \\
C_{3}(z,y) &\approx &\frac{1}{4}\hat{D}_{3}\int_{m}^{\infty }dx\,\left(
x^{2}/m^{2}-1\right) \sin (2y\sqrt{x^{2}-m^{2}})e^{-2ax},  \label{Clarge}
\end{eqnarray}%
where $\nu =0,1,2$. The integrals are expressed in terms of the function $%
K_{\nu }(2m\sqrt{a^{2}+y^{2}})$. By using the corresponding asymptotic
expression for large values of the argument, we find%
\begin{eqnarray}
C_{\nu }(z,y) &\approx &-\sqrt{\pi }am^{3/2}\left[ e^{2mz}+e^{2m(a-z)}\right]
\frac{e^{-2my}}{4y^{7/2}},  \notag \\
C_{3}(z,y) &\approx &-\sqrt{\pi }am^{5/2}\frac{e^{-2my}}{2y^{5/2}},
\label{Clarge2}
\end{eqnarray}%
for $\nu =0,1$ and $C_{2}(z,y)\approx -2myC_{0}(z,y)$. Substituting into (%
\ref{TnuDec2}), we see that for $q>2$ the dominant contribution comes from
the $l=1$ term and for $mr\gg 1$ one gets:%
\begin{eqnarray}
\left\langle T_{\nu }^{\nu }\right\rangle _{\mathrm{b}} &\approx
&\left\langle T_{\nu }^{\nu }\right\rangle _{\mathrm{M}}+\frac{am^{5}}{8\pi
^{3/2}}\left[ e^{2mz}+e^{2m(a-z)}\right] \frac{\cos \left( \pi /q\right)
e^{-2mr\sin (\pi /q)}}{[mr\sin (\pi /q)]^{7/2}},  \notag \\
\left\langle T_{2}^{2}\right\rangle _{\mathrm{b}} &\approx &\left\langle
T_{2}^{2}\right\rangle _{\mathrm{M}}-\frac{am^{5}}{4\pi ^{3/2}}\left[
e^{2mz}+e^{2m(a-z)}\right] \frac{\cos \left( \pi /q\right) e^{-2mr\sin (\pi
/q)}}{[mr\sin (\pi /q)]^{5/2}},  \notag \\
\left\langle T_{3}^{3}\right\rangle _{\mathrm{b}} &\approx &\left\langle
T_{3}^{3}\right\rangle _{\mathrm{M}}+\frac{am^{5}}{4\pi ^{3/2}}\frac{\cos
\left( \pi /q\right) e^{-2mr\sin (\pi /q)}}{[mr\sin (\pi /q)]^{5/2}}.
\label{Tlarger}
\end{eqnarray}%
Note that for the boundary-free part (\ref{EMTsren1}) one has the large
distance asymptotic%
\begin{equation}
\left\langle T_{\nu }^{\nu }\right\rangle _{\mathrm{s}}\approx -\frac{m^{4}}{%
4\pi ^{3/2}}\frac{\cos (\pi /q)e^{-2mr\sin (\pi /q)}}{[mr\sin (\pi /q)]^{5/2}%
},  \label{Tslarge}
\end{equation}%
for $\nu =0,1,3$ and $\left\langle T_{2}^{2}\right\rangle _{\mathrm{s}%
}\approx -3\left\langle T_{0}^{0}\right\rangle _{\mathrm{s}}$. In the case $%
1\leqslant q\leqslant 2$ the suppression of the string-induced contribution
is stronger, by the factor $e^{-2mr}$.

For a massless field, the topological part in the VEV of the energy-momentum
tensor is given by the second term in the right-hand side of (\ref{Tnum0}).
At large distances from the string, $r\gg a$, and for $q>2$ the dominant
contribution to the topological part comes from the term with $l=1$. To the
leading order we find
\begin{equation}
\left\langle T_{\nu }^{\nu }\right\rangle \approx \left\langle T_{\nu }^{\nu
}\right\rangle _{\mathrm{M}}+\frac{\pi A_{\nu }^{(0)}(\pi (r/a)\sin (\pi /q))%
}{4a^{3}re^{\pi (r/a)\sin (\pi /q)}}\cot \left( \pi /q\right) ,
\label{Tlargerm0}
\end{equation}%
where%
\begin{equation}
A_{0}^{(0)}(y)=A_{1}^{(0)}(y)=\frac{1}{y}%
,\;A_{2}^{(0)}(y)=-A_{3}^{(0)}(y)=-1.  \label{A00}
\end{equation}%
For $q\leqslant 2$, in the topological part of (\ref{Tnum0}) the integral
term remains only and the topological part falls off as $e^{-\pi r/a}$. For
a massless field the boundary-free part $\left\langle T_{\nu }^{\nu
}\right\rangle _{\mathrm{s}}$ decays as $1/r^{4}$. Note that in the region
between two plates the decay of the topological part at large distances from
the string is exponential for both the massive and massless cases.

The VEVs $\left\langle T_{\nu }^{\nu }\right\rangle $ for $\nu =0,1,2$
diverge on the plates. This type of surface divergences are well known in
quantum field theory with boundaries and have been investigated for various
bulk and boundary geometries. In the problem under consideration, for points
away from the cosmic string, $r\neq 0$, the divergences are the same as
those for the plates in Minkowski bulk. This means that the part in the VEV
induced by the cosmic string (the topological part), $\left\langle T_{\nu
}^{\nu }\right\rangle _{\mathrm{t}}=\left\langle T_{\nu }^{\nu
}\right\rangle -\left\langle T_{\nu }^{\nu }\right\rangle _{\mathrm{M}}$, is
finite on the plates. This feature could be deduced from general arguments.
Indeed, for points $r\neq 0$, both the local bulk and boundary geometries
for the cosmic string and for Minkowski bulks are the same. The divergences
are determined by the local geometrical characteristics (curvature tensors
for the bulk and boundaries) and, consequently, they are the same as well.
In order to find the leading terms in the asymptotic expansion of $%
\left\langle T_{\nu }^{\nu }\right\rangle $, $\nu =0,1,2$, for points near
the plate at $z=0$, we note that the dominant contribution to the integral
in (\ref{T33Mink}) comes from large values of $x$. One can see that to the
leading order $\left\langle T_{\nu }^{\nu }\right\rangle \approx -m/(12\pi
^{2}z^{3})$. The axial stress is finite on the plates.

\begin{figure}[tbph]
\begin{center}
\epsfig{figure=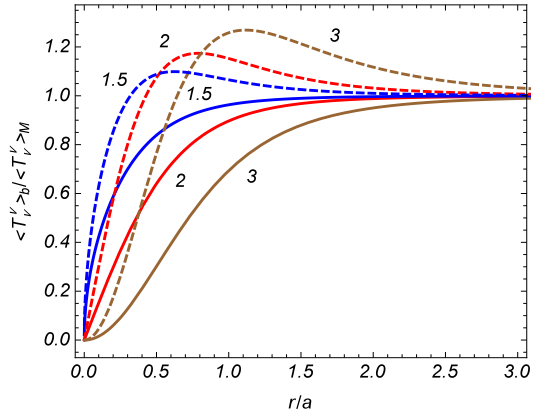,width=7.cm,height=5.5cm}
\end{center}
\caption{The ratio of the plate-induced contributions in the energy density
(full curves) and in the azimuthal stress (dashed curves) to the
corresponding quantities in the Minkowski bulk as functions of the distance
from the string for a massless fermionic field. The numbers near the curves
are the values of the parameter $q$.}
\label{fig2}
\end{figure}

In figure \ref{fig2} we have plotted the ratio of the boundary-induced parts
in the energy density ($\nu =0$, full curves) and the azimuthal stress ($\nu
=2$, dashed curves) to the corresponding quantities for parallel plates in
Minkowski spacetime versus the distance from the cosmic string. The graphs
are plotted for a massless fermionic field and the numbers near the curves
are the values of the parameter $q$.

We could apply the summation formula (\ref{Sumxn}) to the series over $n$ in
the formula (\ref{Tnu}). This gives the following equivalent representation
for the diagonal components of the vacuum energy-momentum tensor (no
summation over $\nu $):%
\begin{eqnarray}
\left\langle T_{\nu }^{\nu }\right\rangle &=&\left\langle T_{\nu }^{\nu
}\right\rangle _{\mathrm{s}}-\frac{q}{\pi ^{2}}\sum_{j}\int_{0}^{\infty
}d\lambda \,\lambda \bar{g}_{qj-1/2}^{(\nu )}(\lambda r)  \notag \\
&&\times \int_{\sqrt{\lambda ^{2}+m^{2}}}^{\infty }dx\,\frac{f^{(\nu )}(x,z)%
}{\frac{x+m}{x-m}e^{2ax}+1}\left( x^{2}-\lambda ^{2}-m^{2}\right) ^{-1/2}.
\label{Tnualt}
\end{eqnarray}%
Here, we have defined the functions%
\begin{equation}
\bar{g}_{\beta }^{(0)}(y)=\bar{g}_{\beta }^{(3)}(y)=J_{\beta
}^{2}(y)+J_{\beta +1}^{2}(y),  \label{gbar}
\end{equation}%
$\bar{g}_{\beta }^{(\nu )}(y)=g_{\beta }^{(\nu )}(y)$ for $\nu =1,2$, and%
\begin{eqnarray}
f^{(0)}(x,z) &=&\left( x^{2}-\lambda ^{2}-m^{2}\right) f(x,z),  \notag \\
f^{(1)}(x,z) &=&f^{(2)}(x,z)=\lambda ^{2}f(x,z),\;f^{(3)}(x,z)=-2x^{2},
\label{f3}
\end{eqnarray}%
with%
\begin{equation}
f(x,z)=2+m\frac{e^{2xz}+e^{2x(a-z)}}{x-m}.  \label{fxz}
\end{equation}%
The second term in the right-hand side of (\ref{Tnualt}) presents the
contribution induced by the boundaries. As it has been shown above, the
radial stress is equal to the energy density. From (\ref{Tnualt}) it follows
that for $r\neq 0$ the boundary-induced contribution is negative to the
energy density and positive for the axial stress. By taking into account
that the boundary-free part in the energy density is negative as well, we
conclude that the total energy density is negative everywhere, $\left\langle
T_{0}^{0}\right\rangle <0$.

We have already shown that the boundary-induced VEVs vanish on the cosmic
string. For points near the string, the contribution of the term $j=1/2$
into the boundary-induced part in (\ref{Tnualt}) dominates and to the
leading order we get (no summation over $\nu $)%
\begin{equation}
\left\langle T_{\nu }^{\nu }\right\rangle \approx \left\langle T_{\nu }^{\nu
}\right\rangle _{\mathrm{s}}-\frac{g_{\nu }\pi ^{-3/2}(r/2)^{q-1}}{%
(q+2)\Gamma (q/2)\Gamma ((q+1)/2)}\int_{m}^{\infty }dx\,\frac{%
(x^{2}-m^{2})^{q/2+1}}{\frac{x+m}{x-m}e^{2ax}+1}f(x,z),  \label{Tnur0}
\end{equation}%
for $\nu =0,1,2$ with%
\begin{equation}
g_{0}=g_{1}=1,\;g_{2}=q,  \label{g012}
\end{equation}%
and%
\begin{equation}
\left\langle T_{3}^{3}\right\rangle \approx \left\langle
T_{3}^{3}\right\rangle _{\mathrm{s}}+\frac{2\pi ^{-3/2}(r/2)^{q-1}}{\Gamma
(q/2)\Gamma ((q+1)/2)}\int_{m}^{\infty }dx\,\frac{x^{2}(x^{2}-m^{2})^{q/2}}{%
\frac{x+m}{x-m}e^{2ax}+1}.  \label{T3r0}
\end{equation}%
In the special case $q=1$ these asymptotics are reduced to the exact results
(\ref{T33Mink}).

\section{Casimir force}

\label{sec:Force}

The Casimir force acting per unit surface of the plate is determined by the
normal stress $\left\langle T_{3}^{3}\right\rangle $. For the corresponding
effective pressure one has: $p=-\left\langle T_{3}^{3}\right\rangle $. The
boundary free part of the pressure is the same on both the sides of the
plate and it does not contribute to the net force. Hence, the force per unit
surface of the plate is determined by the boundary-induced part of the
pressure along the $z$-direction. By taking into account that $\left\langle
T_{3}^{3}\right\rangle _{\mathrm{b}}^{(1)}=0$, one finds%
\begin{eqnarray}
p(r) &=&p_{\mathrm{M}}-\frac{2}{\pi ^{2}}\bigg[%
\sum_{l=1}^{p}(-1)^{l}c_{l}F(rs_{l})+\frac{2q}{\pi }\cos \left( \frac{q\pi }{%
2}\right)   \notag \\
&&\times \int_{0}^{\infty }dx\,\frac{\sinh \left( qx\right) \sinh
(x)F(r\cosh x)}{\cosh (2qx)-\cos (q\pi )}\bigg],  \label{p3}
\end{eqnarray}%
with the function%
\begin{equation}
F(y)=\frac{1}{y}\int_{m}^{\infty }dx\frac{x^{2}\sin (2y\sqrt{x^{2}-m^{2}})}{%
\frac{x+m}{x-m}e^{2ax}+1}.  \label{Fy}
\end{equation}%
Here $p_{\mathrm{M}}=-\left\langle T_{3}^{3}\right\rangle _{\mathrm{M}}$,
with $\left\langle T_{3}^{3}\right\rangle _{\mathrm{M}}$ from (\ref{T33Mink}%
), is the fermionic Casimir pressure for the plates in Minkowski bulk. From
the results of the previous section it follows that for $q>1$ the Casimir
pressure vanishes on the string as $r^{q-1}$. An alternative representation
for the Casimir pressure is obtained by using the formula (\ref{Tnualt}) for
the axial stress:%
\begin{eqnarray}
p(r) &=&-\frac{2q}{\pi ^{2}}\sum_{j}\int_{0}^{\infty }d\lambda \,\lambda %
\left[ J_{qj-1/2}^{2}(\lambda r)+J_{qj+1/2}^{2}(\lambda r)\right]   \notag \\
&&\times \int_{\sqrt{\lambda ^{2}+m^{2}}}^{\infty }dx\,x^{2}\frac{\left(
x^{2}-\lambda ^{2}-m^{2}\right) ^{-1/2}}{\frac{x+m}{x-m}e^{2ax}+1}.
\label{palt}
\end{eqnarray}%
For $r\neq 0$ this pressure is always negative which means that the Casimir
forces are attractive.

At large distances from the string and for a massive field, $mr\gg 1$, from (%
\ref{Tlarger}) for $q>2$ it follows that
\begin{equation}
p(r)\approx p_{\mathrm{M}}-\frac{am^{5}}{4\pi ^{3/2}}\frac{\cos \left( \pi
/q\right) e^{-2mr\sin (\pi /q)}}{[mr\sin (\pi /q)]^{5/2}}.  \label{plarger}
\end{equation}%
For $q<2$ and if $q$ is not too close to 2 one has%
\begin{equation}
p\approx p_{M}+\frac{q^{2}m^{3}\cos \left( q\pi /2\right) e^{-2mr}}{8\pi
^{2}\sin ^{2}(q\pi /2)(mr)^{3}},  \label{plarger1}
\end{equation}%
with the exponential suppression of the topological part.

For a massless field, in (\ref{p3}) one has $p_{\mathrm{M}}=-7\pi
^{2}/(960a^{4})$ and%
\begin{eqnarray}
F(y) &=&-\frac{\pi }{16ay}\partial _{y}^{2}\left[ \frac{a}{\pi y}-\frac{1}{%
\sinh (y\pi /a)}\right]  \notag \\
&=&-\frac{1}{8y^{4}}\left[ 1-\left( \frac{y\pi }{a}\right) ^{3}\frac{1+\sinh
^{2}(y\pi /a)/2}{\sinh ^{3}(y\pi /a)}\right] .  \label{Fym0}
\end{eqnarray}%
At large distances from the string, this leads to the following asymptotic:%
\begin{equation}
p\approx -\frac{7\pi ^{2}}{960a^{4}}\left[ 1+\frac{(q^{2}-1)(q^{2}+17/7)}{%
3\pi ^{4}(r/a)^{4}}\right] ,\;r\gg a.  \label{Plargerm0}
\end{equation}%
In this case one has a power law decay for the topological part. As it has
been shown before, for a massless field and at large distances from the
string the topological part of the axial stress in the region between the
plates is suppressed by the factor $e^{-\pi (r/a)\sin (\pi /q)}$ for $q>2$
and by $e^{-\pi r/a}$ for $q\leqslant 2$. In (\ref{Plargerm0}), the leading
term in the topological part (the second term in the square brackets) comes
from the pressure in the exterior region ($z\leqslant 0$ for the plate at $%
z=0$ and $z\geqslant a$ for the plate at $z=a$). The latter is dominated by
the boundary-free part and induces an attractive force.

Let us denote by $\varepsilon _{\mathrm{t}}$ the topological part in the
vacuum energy per unit surface of the plates:%
\begin{equation}
\varepsilon _{\mathrm{t}}=\frac{1}{\phi _{0}}\int_{0}^{a}dz\int_{0}^{\phi
_{0}}d\phi \,\left\langle T_{0}^{0}\right\rangle _{\mathrm{t}}.
\label{epscs}
\end{equation}%
where the topological contribution to the energy density is obtained from (%
\ref{TnuDec2p}) omitting the $l=0$ term (corresponding to $\left\langle
T_{0}^{0}\right\rangle _{\mathrm{M}}$) in the boundary-induced part (\ref%
{TnuDec2}). Note that $\varepsilon _{\mathrm{t}}$ depends on the radial
coordinate $r$. Integrating by parts, one can see that
\begin{equation}
\partial _{a}\int_{0}^{a}dz\,C_{0}(z,y)=4F(y).  \label{Crel}
\end{equation}%
From here the following relation is obtained between the topological parts
in the vacuum energy and the pressure on the plates:%
\begin{equation}
p_{\mathrm{t}}=-\partial _{a}\varepsilon _{\mathrm{t}},  \label{pepsrel}
\end{equation}%
where $p_{\mathrm{t}}$ is given by the second term in the right-hand side of
(\ref{p3}). This is the standard thermodynamical relation between the energy
and pressure for adiabatic processes. Note that the Minkowkian part $%
\left\langle T_{0}^{0}\right\rangle _{\mathrm{M}}$ diverges on the plates
and in order to obtain the finite vacuum energy an additional
renormalization is required.

\begin{figure}[tbph]
\begin{center}
\begin{tabular}{cc}
\epsfig{figure=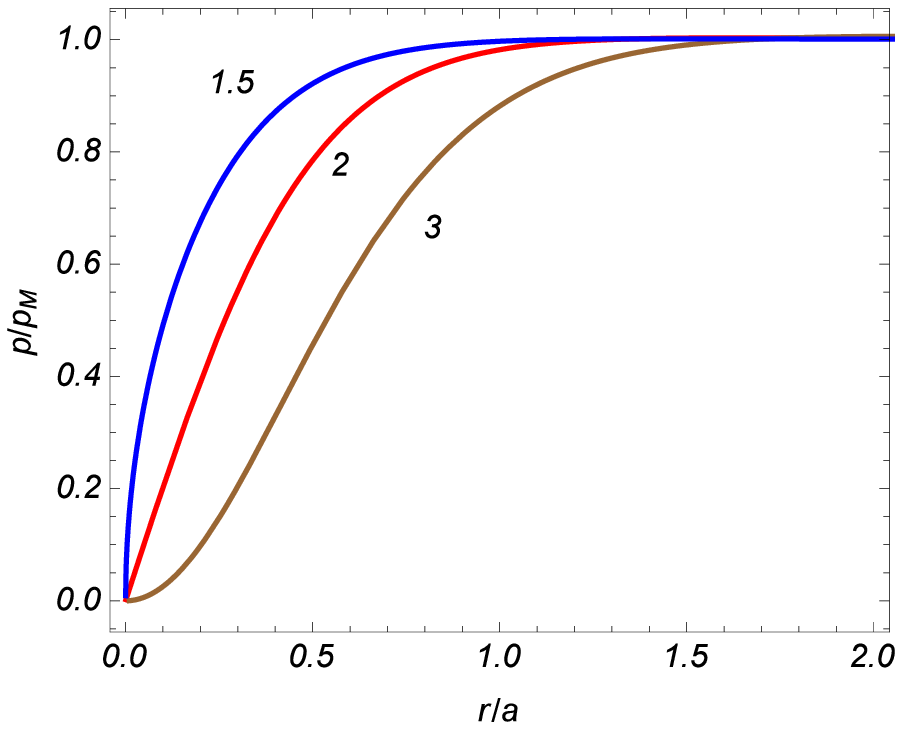,width=7.cm,height=5.5cm} & \quad %
\epsfig{figure=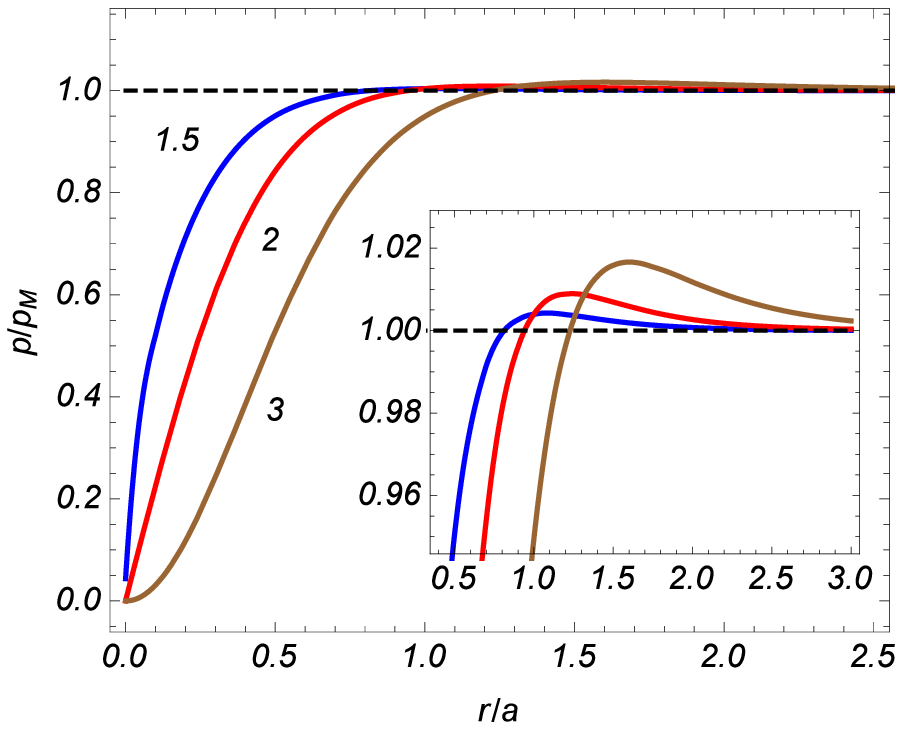,width=7.cm,height=5.5cm}%
\end{tabular}%
\end{center}
\caption{The ratio of the Casimir pressure to the corresponding quantity in
the Minkowski bulk versus the distance from the string. The left panel is
for a massless field and for the right panel $ma=0.5$. The numbers near the
curves are the values of the parameter $q$.}
\label{fig3}
\end{figure}

Figure \ref{fig3} presents the ratio of the Casimir force per unit surface
of the plate to the corresponding quantity in the Minkowski spacetime as a
function of the distance from the cosmic string. The left panel is plotted
for a massless field and for the right panel we have taken $ma=0.5$. In the
latter case for the Minkowskian pressure one has $p_{\mathrm{M}}\approx
-0.0384/a^{4}$. The numbers near the curves are the values of the parameter $%
q$. As is seen from the right panel, the dependence, in general, is not
monotonic.

\section{Conclusion}

\label{sec:Conc}

We have investigated the influence of the conical geometry on the
characteristics of the fermionic vacuum in the region between two plates
perpendicular to the cone axis. This geometry describes an idealized cosmic
string with zero thickness core. On the plates the field obeys the boundary
condition used in bag models to confine the quarks inside the hadrons. Among
the most important local characteristics of the fermionic vacuum are the
fermionic condensate and the expectation value of the energy-momentum
tensor. For the evaluation of the corresponding VEVs we have constructed the
complete set of positive- and negative-energy fermionic modes, given by the
expressions (\ref{psiPos}) and (\ref{psiNeg}). The eigenvalues of the
quantum number, corresponding to the direction along the cosmic string axis,
are quantized by the boundary conditions and are solutions of the
transcendental equation (\ref{EigEq2}). The mode sums of the VEVs contain
series over these eigenvalues. The application of the Abel-Plana type
formula (\ref{Sumxn}) allows to extract the parts in the VEVs for the
geometry of a single plate and to present the second plate induced
contributions in the form for which the explicit knowledge of the
eigenvalues of the axial quantum number is not required.

The fermionic condensate is decomposed into the boundary-free and
boundary-induced contributions (see (\ref{FCdec2})). The boundary-free part
is given by (\ref{FCsren}) and it vanishes for a massless field. For the
boundary-induced part we have derived the expression (\ref{FCC}) with the
function $C(z,y)$ defined as (\ref{Czy}). The corresponding integral is
further evaluated for a massless field with the expression (\ref{Czym0}).
The $l=0$ term in the boundary-induced contribution (\ref{FCC}) corresponds
to the fermion condensate for two parallel plates in Minkowski bulk. The
remaining part is induced by the nontrivial topology of the cosmic string.
We have shown that, for points outside the string core, the topological part
vanishes on the plates as a consequence of the cancellation between the
boundary-free and boundary-induced parts. The Minkowskian part diverges on
the plates with the leading term inversely proportional to the cube of the
distance from the plate. For points away from the plates, the
boundary-induced fermionic condensate vanishes on the string as $r^{q-1}$.
For $q>2$, at large distances from the cosmic string the topological part in
the boundary-induced fermionic condensate decays as $e^{-2mr\sin (\pi /q)}$
for a massive field and as $e^{-2r\sin (\pi /q)/a}$ for a massless field. In
the case $q\leqslant 2$ the decay is stronger, like $e^{-2mr}$ and $%
e^{-2r/a} $ for the massive and massless cases, respectively. An alternative
representation for the boundary-induced contribution to the fermionic
condensate is given by the second term in the right-hand side of (\ref{FCalt}%
). This contribution is negative for points outside the cosmic string. For
points near the cosmic string the fermionic condensate is dominated by the
boundary-free part and is positive. At large distances from the string, the
Minkowskian part dominates and the condensate is negative. Consequently, at
some intermediate distances the fermionic condensate vanishes.

The vacuum energy-momentum tensor is diagonal and the corresponding radial
stress is equal to the energy density. The boundary-induced contributions to
the separate components are given by (\ref{TnuDec2}), where the functions $%
C_{\nu }(z,y)$ are defined by the expressions (\ref{Cnu}) and (\ref{Cnum0})
for massive and massless fields respectively. In the latter case the single
plate parts vanish and the VEV of the energy-momentum tensor in the region
between the plates does not depend on the coordinate $z$. For the axial
stress, that is the case for a massive field as well. Similar to the
fermionic condensate, for points away the plates, the boundary-induced
contributions to the vacuum energy-momentum tensor vanish on the string. At
distances from the cosmic string larger than the Compton wavelength, $mr\gg 1
$, the asymptotic for the topological parts is given by (\ref{Tlarger}) and
they are suppressed by the factor $e^{-2mr\sin (\pi /q)}$ for $q>2$ and by
the factor $e^{-2mr}$ for $q\leqslant 2$. Compared with the azimuthal and
axial stresses, the topological contributions to the energy density and the
radial stress contain an additional suppression factor $1/(mr)$. For a
massless field and at distances $r\gg a$, one has the asymptotic behavior (%
\ref{Tlargerm0}) with the suppression factor $e^{-2r\sin (\pi /q)/a}$ for $%
q>2$ and $e^{-2r/a}$ for $q\leqslant 2$. This behavior for the topological
part is in contrast to that for the boundary-free geometry. In the latter
case the decay of the vacuum energy-momentum tensor is power law, as $1/r^{4}
$. Another representation for the VEV of the energy-momentum tensor in the
region between the boundaries is given by (\ref{Tnualt}) with the last term
been the boundary-induced contribution. Both the boundary-free and
boundary-induced contributions to the energy density are negative.

We have also considered the Casimir force per unit surface of the plates
(the vacuum effective pressure), given by (\ref{p3}). It is decomposed into
the Minkowskian and topological parts, the latter being induced by the
cosmic string. The vacuum pressure on the plates is not homogeneous. It
vanishes at the point where the cosmic string crosses the plates and is
negative at other points. This corresponds to the attractive force between
the plates. The Casimir pressure on the plates, in general, is not a
monotonic function of the distance from the string. We have shown that the
topological contributions in the vacuum energy and the pressure obey the
standard thermodynamical relation (\ref{pepsrel}).

\section*{Acknowledgments}

A.A.S. was supported by the State Committee of Science Ministry of Education
and Science RA, within the frame of Grant No. SCS 15T-1C110, and by the
Armenian National Science and Education Fund (ANSEF) Grant No. hepth-4172.

\appendix

\section{Transformations for the energy-momentum tensor}

\label{sec:Appendix}

Here we describe some details for the presentation of the VEVs (\ref{Tnu})
in the form (\ref{TnuComb}). Let us start with the energy density. The
corresponding mode sum (\ref{modesum}) is rewritten in the form%
\begin{eqnarray}
\left\langle T_{0}^{0}\right\rangle  &=&-\frac{q}{\pi a}\sum_{n=1}^{\infty }%
\frac{h_{0}(x_{n}/a,z)}{1-\sin (2x_{n})/(2x_{n})}  \notag \\
&&\times \sum_{j}\int_{0}^{\infty }d\lambda \,\lambda \omega \left[
J_{qj-1/2}^{2}(\lambda r)+J_{qj+1/2}^{2}(\lambda r)\right] ,  \label{T001}
\end{eqnarray}%
with the notation $h_{0}(k,z)$ defined in accordance with (\ref{hnu}). For
the further transformation we use the integral representation $\omega =-(2/%
\sqrt{\pi })\int_{0}^{\infty }ds\partial _{s^{2}}e^{-\omega ^{2}s^{2}}$.
Integrating over $\lambda $ with the help of (\ref{Jint}) and after the
integration by parts in the integral over $s$, we get%
\begin{eqnarray}
\left\langle T_{0}^{0}\right\rangle  &=&\frac{\pi ^{-3/2}q}{\sqrt{2}r^{3}a}%
\sum_{n=1}^{\infty }\frac{h_{0}(x_{n}/a,z)}{1-\sin (2x_{n})/(2x_{n})}  \notag
\\
&&\times \int_{0}^{\infty }dy\,y^{1/2}e^{-(x_{n}^{2}/a^{2}+m^{2})r^{2}/2y-y}%
\mathcal{I}(q,y).  \label{T002}
\end{eqnarray}%
As the next step we employ the integral representation (\ref{RepIcal}).
After the integration over $y$, for the energy density this gives%
\begin{eqnarray}
\left\langle T_{0}^{0}\right\rangle  &=&-\frac{1}{4\pi ar}\partial _{r}\frac{%
1}{r}\sum_{n=1}^{\infty }\frac{h_{0}(x_{n}/a,z)}{1-\sin (2x_{n})/(2x_{n})}%
\left[ \sideset{}{'}{\sum}_{l=0}^{p}(-1)^{l}\frac{c_{l}}{s_{l}^{3}}%
e^{-2rs_{l}\sqrt{x_{n}^{2}/a^{2}+m^{2}}}\right.   \notag \\
&&\left. +\frac{2q}{\pi }\cos \left( \frac{q\pi }{2}\right) \int_{0}^{\infty
}du\frac{\sinh \left( qu\right) \sinh u}{\cosh (2qu)-\cos (q\pi )}\frac{%
e^{-2r\sqrt{x_{n}^{2}/a^{2}+m^{2}}\cosh u}}{\cosh ^{3}u}\right] .
\label{T003}
\end{eqnarray}%
This representation is in the form (\ref{TnuComb}).

Now let us consider the VEV\ of the azimuthal stress. By taking into account
the relation%
\begin{equation}
J_{\nu }(\lambda r)J_{\nu +1}(\lambda r)=\left( \frac{\nu }{r}-\frac{1}{2}%
\frac{d}{dr}\right) \frac{1}{\lambda }J_{\nu }^{2}(\lambda r),  \label{RelJ}
\end{equation}%
for the Bessel function, the corresponding expression is presented in the
form
\begin{eqnarray}
\left\langle T_{2}^{2}\right\rangle &=&-\frac{2q^{2}}{\pi ar}%
\sum_{n=1}^{\infty }\frac{h_{2}(x_{n}/a,z)}{1-\sin (2x_{n})/(2x_{n})}  \notag
\\
&&\times \sum_{j}j\left( \frac{qj-1/2}{r}-\frac{1}{2}\partial _{r}\right)
\int_{0}^{\infty }d\lambda \,\frac{\lambda }{\omega }J_{qj-1/2}^{2}(\lambda
r).  \label{T221}
\end{eqnarray}%
By using the integral representation (\ref{IntRep}), after the integration
over $\lambda $, the VEV is rewritten as%
\begin{eqnarray}
\left\langle T_{2}^{2}\right\rangle &=&\frac{4q^{2}r^{-3}}{\left( 2\pi
\right) ^{3/2}a}\sum_{n=1}^{\infty }\frac{h_{2}(x_{n}/a,z)}{1-\sin
(2x_{n})/(2x_{n})}\sum_{j}j\int_{0}^{\infty }dy  \notag \\
&&\times y^{-1/2}e^{-(k^{2}+m^{2})r^{2}/2y}\left( qj-1/2-y\partial
_{y}\right) e^{-y}I_{qj-1/2}(y).  \label{T222}
\end{eqnarray}%
Now, with the help of the relation%
\begin{equation}
\left( qj-1/2-y\partial _{y}\right) e^{-y}I_{qj-1/2}(y)=\frac{y}{qj}%
e^{-y}\left( y\partial _{y}-y+1/2\right) \left[ I_{qj-1/2}(y)+I_{qj+1/2}(y)%
\right] ,  \label{RelI}
\end{equation}%
the series over $j$ is expressed in terms of the function $\mathcal{I}(q,y)$%
:
\begin{eqnarray}
\left\langle T_{2}^{2}\right\rangle &=&\frac{2qr^{-3}}{\left( 2\pi \right)
^{3/2}a}\sum_{n=1}^{\infty }\frac{h_{2}(x_{n}/a,z)}{1-\sin (2x_{n})/(2x_{n})}%
\int_{0}^{\infty }dy  \notag \\
&&\times e^{-(k^{2}+m^{2})r^{2}/2y-y}y^{1/2}\left( y\partial
_{y}-y+1/2\right) \mathcal{I}(q,y).  \label{T223}
\end{eqnarray}%
By making use of the representation (\ref{RepIcal}) we get%
\begin{eqnarray}
\left\langle T_{2}^{2}\right\rangle &=&\frac{-1}{2\pi ar^{3}}(1-r\partial
_{r}+\frac{1}{2}r^{2}\partial _{r}^{2})\sum_{n=1}^{\infty }\frac{%
h_{2}(x_{n}/a,z)}{1-\sin (2x_{n})/(2x_{n})}\left[ \sideset{}{'}{\sum}%
_{l=0}^{p}(-1)^{l}\frac{c_{l}}{s_{l}^{3}}e^{-2rs_{l}\sqrt{%
x_{n}^{2}/a^{2}+m^{2}}}\right.  \notag \\
&&\left. +\frac{2q}{\pi }\cos \left( \frac{q\pi }{2}\right) \int_{0}^{\infty
}dx\frac{\sinh \left( qx\right) \sinh x}{\cosh (2qx)-\cos (q\pi )}\frac{%
e^{-2r\sqrt{x_{n}^{2}/a^{2}+m^{2}}\cosh x}}{\cosh ^{3}x}\right] .
\label{T224}
\end{eqnarray}

For the axial stress the mode sum is reduced to%
\begin{equation}
\left\langle T_{3}^{3}\right\rangle =\frac{q}{\pi a^{3}}\sum_{j}\int_{0}^{%
\infty }d\lambda \,\lambda \sum_{n=1}^{\infty }\frac{x_{n}^{2}}{\omega _{n}}%
\frac{J_{qj-1/2}^{2}(\lambda r)+J_{qj+1/2}^{2}(\lambda r)}{1-\sin
(2x_{n})/(2x_{n})}.  \label{T331}
\end{equation}%
With the help of (\ref{IntRep}), after the integration over $\lambda $, we
find%
\begin{equation}
\left\langle T_{3}^{3}\right\rangle =\frac{qa^{-3}}{\sqrt{2}\pi ^{3/2}r}%
\sum_{n=1}^{\infty }\frac{x_{n}^{2}}{1-\sin (2x_{n})/(2x_{n})}%
\int_{0}^{\infty }dyy^{-1/2}e^{-\left( m^{2}+k^{2}\right) r^{2}/2y-y}%
\mathcal{I}(q,y).  \label{T332}
\end{equation}%
By taking into account (\ref{RepIcal}) and integrating over $y$ one gets the
representation%
\begin{eqnarray}
\left\langle T_{3}^{3}\right\rangle &=&\frac{1}{\pi a^{3}r}%
\sum_{n=1}^{\infty }\frac{x_{n}^{2}}{1-\sin (2x_{n})/(2x_{n})}\left[ %
\sideset{}{'}{\sum}_{l=0}^{p}\frac{(-1)^{l}\cot \left( \pi l/q\right) }{%
e^{2rs_{l}\sqrt{x_{n}^{2}/a^{2}+m^{2}}}}\right.  \notag \\
&&\left. +\frac{2q}{\pi }\cos \left( \frac{q\pi }{2}\right) \int_{0}^{\infty
}dx\frac{\sinh \left( qx\right) \tanh x}{\cosh (2qx)-\cos (q\pi )}e^{-2r%
\sqrt{x_{n}^{2}/a^{2}+m^{2}}\cosh x}\right] .  \label{T333}
\end{eqnarray}%
The expression for the radial stress is obtained from the trace relation (%
\ref{TrRel}).

\end{document}